%% file: main.tex
\documentclass[%
 aip,
 amsmath,amssymb,
 reprint
]{revtex4-1}
\usepackage[utf8]{inputenc}
\usepackage[T1]{fontenc}
\usepackage{mathptmx}
\usepackage{etoolbox}

\usepackage{graphicx}% Include figure files
\usepackage{dcolumn}% Align table columns on decimal point
\usepackage{bm}

\makeatletter
\def\@email#1#2{%
 \endgroup
 \patchcmd{\titleblock@produce}
  {\frontmatter@RRAPformat}
  {\frontmatter@RRAPformat{\produce@RRAP{*#1\href{mailto:#2}{#2}}}\frontmatter@RRAPformat}
  {}{}
}%
\makeatother
\preprint{AIP/123-QED}

\usepackage{braket}
\usepackage{mathtools}
\usepackage{ifthen}
\usepackage{listings}
\usepackage{pythonhighlight}
\usepackage{appendix}
\input{defs}

\usepackage{amsthm}

\theoremstyle{definition}

\theoremstyle{remark}

\usepackage{multirow}

\usepackage{tikz}
\usetikzlibrary{arrows.meta}

\usepackage{mathtools}

\begin{document}

\title[Expanding Density-Correlation Machine Learning Representations for Anisotropic Coarse-Grained Particles]{Expanding Density-Correlation Machine Learning Representations for Anisotropic Coarse-Grained Particles}
\author{Arthur Lin}
\affiliation{Department of Chemical and Biological Engineering, University of Wisconsin, Madison, WI, USA}
\author{Kevin K. Huguenin-Dumittan}
\affiliation{Laboratory of Computational Science and Modeling, École Polytechnique Fédérale de Lausanne, Lausanne, Switzerland}
\author{Yong-Cheol Cho}
\affiliation{Department of Chemical and Biological Engineering, University of Wisconsin, Madison, WI, USA}
\affiliation{Department of Computer Science and Engineering, University of Wisconsin, Madison, WI, USA}
\author{Jigyasa Nigam}
\affiliation{Laboratory of Computational Science and Modeling, École Polytechnique Fédérale de Lausanne, Lausanne, Switzerland}
\author{Rose K. Cersonsky}
\affiliation{Department of Chemical and Biological Engineering, University of Wisconsin, Madison, WI, USA}
\date{\today}

\begin{abstract}
Physics-based, atom-centered machine learning (ML) representations have been instrumental to the effective integration of ML within the atomistic simulation community. Many of these representations build off the idea of atoms as having spherical, or isotropic, interactions. In many communities, there is often a need to represent groups of atoms, either to increase the computational efficiency of simulation via coarse-graining or to understand molecular influences on system behavior. In such cases, atom-centered representations will have limited utility, as groups of atoms may not be well-approximated as spheres. In this work, we extend the popular Smooth Overlap of Atomic Positions (SOAP) ML representation for systems consisting of non-spherical anisotropic particles or clusters of atoms. We show the power of this anisotropic extension of SOAP, which we deem \AniSOAP, in accurately characterizing liquid crystal systems and predicting the energetics of Gay-Berne ellipsoids and coarse-grained benzene crystals. With our study of these prototypical anisotropic systems, we derive fundamental insights on how molecular shape influences mesoscale behavior and explain how to reincorporate important atom-atom interactions typically not captured by coarse-grained models. Moving forward, we propose \AniSOAP as a flexible, unified framework for coarse-graining in complex, multiscale simulation.
\end{abstract}

\maketitle

\section{Introduction}

In understanding molecular interactions and modeling their resultant behavior, it is very often a worthwhile endeavor to group (i.e. coarse-grain) one or more atoms as a theoretical ``unit'' or particle. This choice can be practical; often, the time- and length-scales necessary to simulate molecular processes limit our ability to simulate with all-atom resolution. Conversely, choosing particle-based, rather than atom-based computational approaches can also be a scientific choice; when we selectively limit the degrees of freedom within our systems, we can identify factors that are causal to molecular behavior or phenomena. Thus, these simplified coarse-grained simulation approaches serve as both a tool and a lens with which to study chemical systems.

Similarly, machine learning (ML) methods have emerged as a powerful tool for scientific inquiry, with the ability to elucidate new patterns within or relationships between chemical spaces and observed properties, often in order to predict the properties of unseen systems. 
So, how do we incorporate the idea of atom grouping in the context of machine learning? Many approaches consider our configurational space as a manifold and use a variety of machine learning architectures (variational\cite{monroe_learning_2022, monroe_systematic_2022} and hierarchical\cite{costa_ophiuchus_2023} auto-encoders, different forms of neural networks\cite{majewski_machine_2023, ruza_temperature-transferable_2020, husic_coarse_2020, wang_machine_2019, zhang_deepcg_2018}, ensemble learning \cite{wang_ensemble_2020}) to determine a latent space in which this grouping is implicitly embedded that minimizes a chosen fitness function.
Despite high performance and generalizability, these end-to-end models are often limited by data requirements or in fundamental analyses by their lack of intrinsic interpretability.

Another approach within the machine-learning community is so-called \emph{feature-forward} modeling, wherein one explicitly transforms the raw chemical data into numerical ``features'' that reflect the underlying physics or chemistry of interest prior to applying ML methods. While both end-to-end and feature-forward approaches have merits and overlaps, the latter method is often noted for its interpretability and comparable performance using shallower ML architectures\cite{de_comparing_2016, deringer_gaussian_2021, musil_physics-inspired_2021}. Within this umbrella of approaches, there are many ways to encode the raw chemical data into features, and the suitable choice depends entirely on the scientific context. For cheminformatics, wherein we are often looking to compare different chemistries or identify the role of specific functional groups, string-based featurizations such as SMILES\cite{weininger_smiles_1988} or SELFIES\cite{krenn_self-referencing_2020} are popular, as they encode important parameters such as present functional groups and connectivity and can be parsed using natural language processing (NLP) models and other text-based technologies. However, in thermodynamic contexts where the chemistry and connectivity remain largely unchanged, such as in molecular simulation, it is more typical to use configuration-dependent features\cite{musil_physics-inspired_2021, wigh_review_2022}, including Behler-Parinello symmetry functions\cite{behler_atom-centered_2011}, smooth overlap of atomic positions (SOAP)\cite{bartok_representing_2013}, and molecular graphs \cite{wu_moleculenet_2018}.

So, for coarse-grained systems, how do we apply a feature-forward approach?
A large challenge is that groups of atoms, hereon referred to as ``particles'', are non-isotropic, and so methods based on atomistic ML will fail to capture the anisotropy of interparticle interactions. %Recent work has extended the BPSF formalism for anisotropic particles\cite{campos-villalobos_machine-learning_2022} by including an orientational alignment parameter; however, this is limited to systems of one particle geometry.
Frameworks based on density expansions (\eg SOAP\cite{bartok_representing_2013}, NICE\cite{nigam_recursive_2020}) present a compelling avenue for expansion, given that they can putatively be made flexible to \textit{any} density expansion, even anisotropic density fields or hard particle volumes. Furthermore, by extending these frameworks to isolate molecule-level interactions within atom-atom site potentials, we gain the ability to combine or compare representations across multiple scales.

Here, we propose and demonstrate the first such anisotropic expansion of symmetrized density-based frameworks for ML representations by taking the popular SOAP (Smooth Overlap of Atomic Positions) formalism and demonstrating its expansion to simple anisotropic bodies, in a representation we deem \anisoap, which can be read as either ``The Smooth Overlap of Anisotropic Particles'' or ``Anisotropic-SOAP''. While here we demonstrate the expansion for multivariate Gaussian densities, similar expansions can be made for arbitrary anisotropic density fields. 

In Section~\ref{sec:theory}, we provide the underlying mathematical theory behind this expansion, including how deliberate selection of basis sets enables analytical evaluation of the expansion coefficients. 
In Section~\ref{sec:results}, we demonstrate three performance-defining case studies for its usage across multiple simulation length scales and materials systems.
We start by analyzing two classic cases of explicitly ellipsoidal particles: liquid crystalline configurations and those governed by the Gay-Berne interaction potential. 
In doing so, we show the conceptual overlap and divergence of \anisoap from traditional mesoscale order parameters, as well as its ability to generalize to supervised tasks.
Then, we analyze a set of systems that is only \textit{implicitly} ellipsoidal -- benzene molecules arranged in stable and unstable crystalline configurations, and show that \anisoap can be used to provide molecule-level approximations to first-principles energetics. 
We then couple \anisoap with the traditional SOAP formalism to demonstrate how to combine representations across multiple scales. The corresponding open-source code, \AniSOAP, is available at \url{github.com/cersonsky-lab/anisoap}.

\section{Theory}
\label{sec:theory}

\begin{table}
\begin{tabular}{|p{0.17\linewidth}|p{0.83\linewidth}|}
\hline
\multicolumn{2}{|c|}{General Variables} \\
\hline
$\br$  & a vector in Cartesian space \\
\hline
$r$  & magnitude of vector $\br$ \\
\hline
$\hat{r}$  & direction of vector $\br$ \\
\hline
$\brij$ & the vector between point $i$ and point $j$ \\
\hline 
$\rot$ & a 3x3 rotation matrix\\
\hline 
$\bx$ & a feature vector for one configuration\\
\hline 
$\bX$ & a matrix containing, as rows, feature vectors for multiple configurations\\
\hline 
\hline
\multicolumn{2}{|c|}{Density Expansions} \\
\hline
$g(\br)$ & a potentially anisotropic density function centered at $\br$\\
\hline 
$\rho_{ij}$ & the contribution of the density located at the position of point $j$ to the density expansion of point $i$\\
\hline
$n, l, m$ & indices for radial bases and spherical harmonics\\
\hline
$R_{nl}$ & a basis function indexed by $n$ and $l$\\
\hline
$\sigma_\text{GTO}$ & the width of a Gaussian-type orbital (GTO) basis\\
\hline
$Y_l^m$ & the $m^{th}$ component spherical harmonic of order $l$\\
\hline
$\mathcal{D}_{mm'}^{l}$ & the Wigner matrix used to rotate spherical harmonics\\
\hline
\hline
\multicolumn{2}{|c|}{Multivariate Gaussians (MVG)} \\
\hline
$\sigma_1$, $\sigma_2$, $\sigma_3$ & the three principal components of an MVG\\
\hline
$\bD$ & the principal axis decomposition of an MVG, where $\bD \equiv \mathrm{diag}\left(\frac{1}{\sigma_1^2}, \frac{1}{\sigma_2^2}, \frac{1}{\sigma_3^2}\right)$\\
\hline
$\mathbf{A}$ & the rotated principal axis decomposition of an MVG, where $\mathbf{A}\equiv \rot \bD \rot^T$\\
\hline
$\bS$ & the Gay-Berne analog to $\bD$, where $\bS\equiv\bD^{-2}$\\
\hline
\end{tabular}
\caption{\textbf{Notation Guide.} 
Throughout the text, we adopt the notation typical of the atom-centered symmetrized representation community as detailed in the table above.}
\end{table}

In the traditional formalism for SOAP and related representations, we treat each atom as a localized isotropic field in three dimensions and construct an ``atom-density" by summing over all neighboring atoms within a spherical shell. The atom-density is usually written on a basis of radial and angular functions, the latter of which are chosen to be spherical harmonics to make the atom-density more easily amenable to rotational symmetrization.. To represent the many-body nature of these atoms, we can then compute n-body correlations of these density expansions or introduce message-passing into the featurization formalism\cite{nigam_unified_2022, batatia_mace_2022}.

\subsection{The Effect of Symmetry Breaking in Density Frameworks}
Using the braket notation prescribed in Ref. \citenum{musil_efficient_2021}, the contribution of an atom $j$ to center atom $i$'s expansion coefficients are given by
\begin{align}
    \pnlm = \int_{\relc} \gij{} \rnl{} \ylm{} d^3\br.
    \label{eq:coeff_base}
\end{align}
where $\gij{\br}$ is a localized function (usually a gaussian), $\rnl{}$ is a radial basis function, and $\ylm{}$ are spherical harmonics. For a general $\gij{\br}$ and $\rnl{}$, there is no hope to evaluate this analytically. There is not even a general way to evaluate a general one-dimensional integral $\int f(x) \mathrm{d}x,$ which is why, after all, there are so many books on integral tables.

If we use any arbitrary density $g(\br)$, e.g. to model more closely the shape of rigid molecules, nanoparticles, or arbitrary bodies, we are forced to use numerical integration, which could be done using Lebdev grids. A fully numerical implementation would also provide us with complete freedom regarding the choice of the radial basis function $R_{nl}$, which would allow us to choose the basis based on nice mathematical properties like the Laplacian eigenstate (LE) basis\cite{bigi_smooth_2022}. 
The main downside of the numerical approach, at least for sufficiently general densities $g(\br)$, is the inevitable and potentially severe numerical cost-accuracy trade-off. 
The necessary inaccuracies introduced by numerical integration may negate the fidelity of the anisotropic density field, thus, we would therefore like to examine the possibility of a fully analytical approach for the evaluation of these coefficients. 

For isotropic representations, it is traditional to do an implicit transformation to align $\brij$ onto the $z-axis$ of our coordinate system, eliminating any dependence of $g(\br - \brij)$ on the angular integrands.
This simplifies \eqref{eq:coeff_base}, even for complicated bases, to
\begin{align}
    \pnlm = F(r_{ij}) Y_l^m(\hat{r}_{ij}),
    \label{eq:dens_exp}
\end{align}
where $F(r_{ij})$ is some function that contains the complete dependence on the distance between the two atoms. The angular dependence is fully captured in the spherical harmonic factor, allowing us to precompute $F(\brij)$ on the full relevant interval for efficient spline evaluations\cite{musil_efficient_2021}. 

For non-isotropic densities, however, this is no longer the case. Even taking the relatively simple case of a multivariate Gaussian, we cannot decouple our radial (\eg distance) and angular components, as the Gaussian requires a transformation into the three principal components $\sigma_1, \sigma_2, \sigma_3$ and corresponding axes. In order to maintain consistency with \eqref{eq:coeff_base}, we replace all instances of $\br$ with $\rot\rot^T\br$, where $\rot$ is our rotation matrix into this component axis. We then simplify by changing our axes of integration from $\br$ to $\br'\equiv \rot^T\br$ and noting that $r=\rot\rot^Tr$ to obtain $$\pnlm = \int_{\relc}\mathrm{d}^3(\br')~\gij{\rot\br' - \rot\brij'}\rnl{r}\ylm{\rot\hat{r}'}.$$

To perform the rotation of our spherical harmonics, one commonly sums over Wigner-D matrices\cite{wigner_gruppentheorie_1931} to obtain 
\begin{widetext}
\begin{align}
\pnlm & = \sum_{m'=-l}^l \wigD^l_{mm'}(\rot) \int_{\relc}\mathrm{d}^3(\br')~\gij{\rot\left(\br' - \brij'\right)} \rnl{r'} \ylmp{\hat{r}'}
\end{align}
\end{widetext}

With this step, we can reduce our calculation in the special case in which the matrix $\wigD^l_{mm'}(\rot)$ is diagonal, which proves useful for analytical evaluation techniques. We thus focus on the factor
\begin{align}
    \int_{\relc}\mathrm{d}^3\br'~\gij{\rot\left(\br' - \brij'\right)} \rnl{r'} \ylmp{\hat{r'}}.
    \label{eq:coefficient_general}
\end{align}
which will be the focus of our discussion from hereon.
% and instead need to choose particularly ``nice'' functions that are convenient to integrate. I believe that if we have any hope of finding a nice expression, one good candidate is to
% \begin{itemize}
%     \item choose multivariate Gaussian densities
%     \item use either a monomial ($r^{l+2n}$) or a slightly modified version of the GTO basis $(r^{l+2n}e^{-r^2})$
% \end{itemize}

\subsection{Multivariate Gaussian Densities}
The simplest anisotropic function that can be put in place of $g(r)$ in \eqref{eq:coefficient_general} is the multivariate Gaussian density (hereon MVG). The goal of this subsection is to show that by choosing $g(\br)$ to be an MVG and $\rnl{}$ to be of the monomial, GTO, or STO form, we can reduce the evaluation of the expansion coefficient to the evaluation of integrals of the form
\begin{equation}
\begin{split}
    \pnlm =  \int_{\relc}%\mathrm{d}^3r
    e^{-\frac{1}{2}(\br-\brij)^T\bD(\br-\brij)}\mathrm{p}(\br),
    \label{coefficient_general}
\end{split}
\end{equation}
where $\mathrm{p}(\br)$ is some polynomial expression in $\br \equiv (x,y,z)$ and $\bD$ is some $3\times 3$ diagonal matrix.

% \paragraph*{Definitions}
% We begin by introducing the Gaussian density in section 1.2.1, followed by a first simplification of the general integral for general Gaussian densities in section 1.2.2. Section 1.2.3 then discusses the two choices for the radial basis. The next subsection, 1.3, deals with how to evaluate the above expression in practice.

Consider the three-dimensional Gaussian defined by
\begin{align}
    g:\relc & \rightarrow \real \\
    \br & \rightarrow g(\br) =  \exp\left(-\frac{1}{2}\br^T\bA \br \right),
    \label{eq:gauss_A}
\end{align}
where $A$ is a symmetric and (strictly) positive definite $3 \times 3$ matrix. Any matrix satisfying these two conditions can be orthogonally diagonalized, also called the principal axis decomposition, allowing us to write
\begin{align}
    \bA = \rot \bD \rot^T,
\end{align}
where $\rot \in \mathrm{SO}(3)$ is a rotation matrix that specifies the three principal axes and 
\begin{align}
    \bD \equiv \mathrm{diag}\left(\frac{1}{\sigma_1^2}, \frac{1}{\sigma_2^2}, \frac{1}{\sigma_3^2}\right) = \begin{pmatrix} \frac{1}{\sigma_1^2} & & \\ & \frac{1}{\sigma_2^2} & \\ & & \frac{1}{\sigma_3^2}
    \end{pmatrix}
    \label{eq:dilmat}
\end{align}
is a diagonal matrix containing the widths $\sigma_j$ of the Gaussian along the three principal directions. 
With this decomposition, we can write
\begin{align}
    g(\rot\br') & = \exp\left(-\frac{1}{2}\br'^T\rot^T\rot D\rot^T\rot \br' \right) \\
    % & = \exp\left(-\frac{1}{2}(\rot^T\br)^TD(\rot^T\br) \right) \\
    & =: \exp\left(-\frac{1}{2}\br'^T\bD\br' \right),
\label{eq:gauss_D}
\end{align}
where $\br' = \rot^T\br$ are the coordinates with respect to the principal axes.

% \subsubsection{Spherical Expansion Coefficients}

Putting the equation for our MVG into \eqref{eq:coefficient_general}, we get
\begin{align}%\label{coefficient_general}
    \int_{\relc}\mathrm{d}^3\br' \exp\left(-\frac{1}{2}(\br'-\brij')^T\bD(\br'-\brij') \right) \rnl{r'} \ylmp{\hat{r'}}.
    \label{eq:coefficient_multi}
\end{align}

We have now set the stage for the evaluation of the general coefficient. Our goal is to choose a suitable radial basis that would allow us to evaluate the coefficients analytically. One possibility is to use certain polynomial bases, which is motivated by the fact that for any $n=0,1,2,\dots$, we have
\begin{align}
    \int_{\real}\mathrm{d}x x^n e^{-\frac{x^2}{2\sigma^2}} = \begin{cases}
    \left(2\sigma^2\right)^{\frac{n+1}{2}}\Gamma\left(\frac{n+1}{2}\right) & n \quad \mathrm{even} \\ 0 & n \quad \mathrm{odd} \end{cases}
\end{align}
where $\Gamma$ is the gamma function. % Note that using $n' = \frac{n+1}{2}$, we can more conveniently write this as
% \begin{align}
%     \int_{\real}\mathrm{d}x x^n e^{-\frac{x^2}{2\sigma^2}} =  \left(2\sigma^2\right)^{n'}\Gamma(n')
% \end{align}
% {\color{red} exponent of x inconsistent wrt n'}
% for even $n$. This easily generalizes to (diagonalized) three dimensional Gaussians. 
% For instance, for even $n_1,n_2,n_3$, we get \eqref{eq:wide_gamma}.
% \begin{widetext}
% \begin{equation}
%     \int_{\relc}\mathrm{d}^3r x^{n_1}y^{n_2}z^{n_3}e^{-\frac{1}{2}\left(\frac{x^2}{\sigma_x^2} + \frac{y^2}{\sigma_y^2} + \frac{z^2}{\sigma_z^2} \right)}
%     = \left(2\sigma_x^2\right)^{\frac{n_1 + 1}{2}}\Gamma\left(\frac{n_1 + 1}{2}\right)
%     \left(2\sigma_y^2\right)^{n_2'}\Gamma(n_2')
%     \left(2\sigma_z^2\right)^{n_3'}\Gamma(n_3')
%     \label{eq:wide_gamma}
% \end{equation}
% \end{widetext}
If we could reduce the evaluation of the integral \eqref{coefficient_general} to such (multivariate) polynomials, this might provide us with analytical expressions for the expansion coefficients. Thus, following \eqref{eq:coefficient_multi}, we require that $\rnl{} \ylm{}$ is a polynomial. 
\paragraph*{Integrability of Monomial Basis}
We will first show that when $\rnl{}  = r^{l+2n}$, the expansion coefficients can be evaluated analytically. First, we separate $\rnl{}$ into two factors, where 
\begin{equation}
r^{l+2n}  = (r^2)^n \cdot r^l
\end{equation}
Given that $r^l\ylm{}$ is a polynomial in $(x,y,z)$, as is $r^2 = x^2+y^2+z^2$ and thus, so is $(r^2)^n$, the monomial basis can be analytically integrated.

% This result can be used to generate an infinite number of radial basis functions:
% \begin{corollary}
%     For any $(n,l,m)$, 
%     \begin{align}
%         r^{l+2n}\ylm{}
%     \end{align}
%     is a polynomial in the variables $\bX,y,z$.
% \end{corollary}
% \begin{proof}
%     We can rewrite the expression as
%     \begin{align}
%         r^{l+2n}\ylm{}  = (r^2)^n \cdot r^l\ylm{}
%     \end{align}
% \end{proof}
% %
% By the previous proposition, the second factor is a polynomial in $(x,y,z)$. For the first factor, note that $r^2 = x^2+y^2+z^2$ is a polynomial, and thus, so is $(r^2)^n$.

% \begin{theorem}[Integrability of Monomial Basis]
%     Let $\rnl{}  = r^{l+2n}$. Then, the expansion coefficients can be evaluated analytically.
% \end{theorem}
% \begin{proof}
%     Follows immediately by plugging in this basis into \eqref{coefficient_general} and using the previous corollary.
% \end{proof}

\paragraph*{Integrability of GTO Basis}
With one extra step, we can also show that a suitable modification of the GTO basis,  $\rnl{}  = r^{l+2n}e^{-\frac{r^2}{2\sigma^2}}$,  provides an equally well-suited basis. . 
By separating out all the exponential and polynomial factors, we can write the integrand as
\begin{equation}
\begin{split}
    &e^{-\frac{1}{2}(\br'-\brij')^TD(\br'-\brij')} \rnl{r'} \ylm{\hat{r'}} \\
     = &e^{-\frac{1}{2}(\br'-\brij')^TD(\br'-\brij')} r^{l+2n}e^{-\frac{r^2}{2\sigma^2}}\ylm{} \\
     = &e^{-\frac{1}{2}(\br'-\brij')^TD(\br'-\brij') -\frac{r^2}{2\sigma^2}}r^{l+2n}\ylm{} \\
     = &e^{-\frac{1}{2}(\br'-\brij')^TD(\br'-\brij') -\frac{r^2}{2\sigma^2}}\mathrm{p}(x,y,z) \\
     = &e^{-\mathrm{quadratic}(\br)}\mathrm{p}(x,y,z)
\end{split}
\end{equation}
We can see that this is again an exponential of a (now different) quadratic form multiplied by a polynomial in $(x,y,z)$, which can be integrated analytically.

Concluding this section, we can see that for both choices of the basis, the final expression we need to evaluate is of the form
\begin{align}\label{coefficient_essence}
    \int_{\relc}\mathrm{d}^3\br e^{-\frac{1}{2}(\br-\brij)^T\bD(\br-\brij)} r^{2n}\rlm{},
\end{align}
with $\rlm{} = \br^l\ylm{}$ and where the diagonal matrix $\bD$ is either the one directly obtained from the principal components of the Gaussian density (monomial basis), or the modified version (GTO basis). Similar arguments can be made for Slater-type orbitals (STO), which follow a similar form to the GTO basis. From this expansion, we can then calculate n-body correlations for a given particle, which for the 3-body term is

\begin{gather}
    \braket{nl; n'l' | \rho_i^{\nu=2}} = \sum_m \braket{nlm | \rho_i}\braket{n'l'm | \rho_i}
    \label{eq:3b}
\end{gather}
where $\braket{nlm | \rho_i} = \sum_j \braket{nlm|\rho_{ij}}.$
In Appendix.~\ref{si:evaluating}, we have included a discussion of the practical steps toward computing the expansion coefficient, including Löwdin symmetric orthogonalization\cite{lowdin_nonorthogonality_1970} (discussed in Appendix.~\ref{si:normalization}) to impose orthonormality within the basis set.

\section{Results and Discussion}
\label{sec:results}

Even for general point clouds, determining the optimal anisotropic analog is non-trivial, and the focus of ongoing work\cite{satorras_en_2022}. Thus, we choose to demonstrate the efficacy of \AniSOAP on systems where the choice of an ellipsoidal proxy is trivial, being both historically founded and well-defined. We ground our expansion in the rich history of identifying causal mechanisms through anisotropic proxy particles\cite{zhang_self-assembly_2005}, as many complex behaviors within molecular systems can be explained by analyzing the steric interactions of their molecular volumes\cite{glotzer_anisotropy_2007, cersonsky_pressure-tunable_2018, cersonsky_relevance_2018,  keys_characterizing_2011, van_anders_understanding_2014}. Future studies will focus on optimizing \anisoap for less trivial cases and further varieties of molecular anisotropy. 

\begin{figure*}[!ht]
    \centering
    \includegraphics[width=\linewidth]{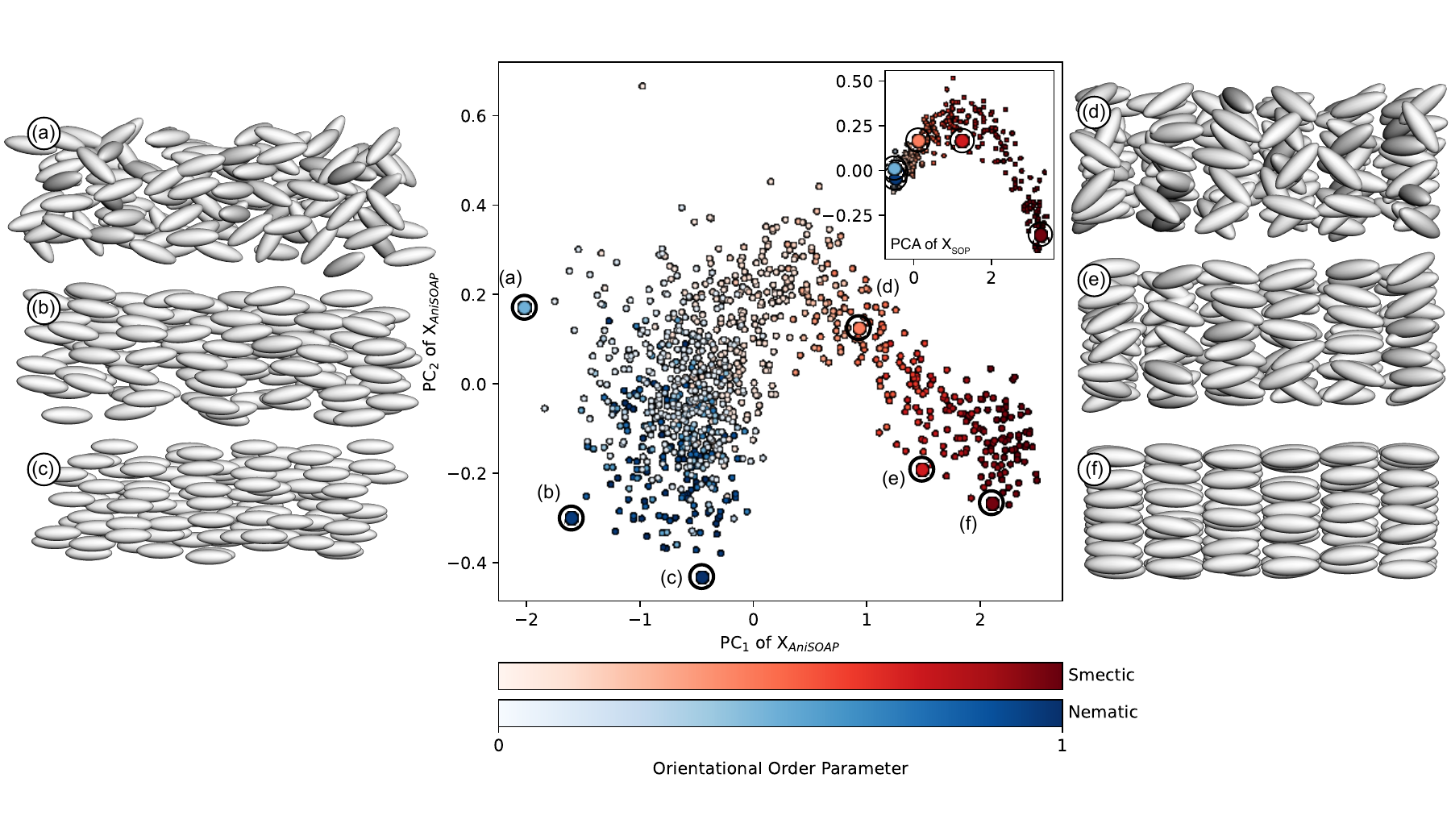}
    \caption{\textbf{The feature space of \anisoap in the context of LC phases.} 
    The middle plot shows the first two components of the structure-average \anisoap vectors for 2000 generated liquid crystal configurations showing either no translational order (nematic phases, blue), planar stacking (smectic phases, red), and varying degrees of orientation order (color saturation, with white being little to no order). The inset shows the first two principal components of the popular Steinhardt Order parameters, which clearly delineate the smectic (red) phases, but in these and all following components, do not separate the different nematic (blue) phases. Representative snapshots are shown to the left and right corresponding to labeled points on the plots.
    }
    \label{fig:lc_pca}
\end{figure*}
\subsection{Ellipsoids in Liquid Crystals}
\label{sec:lc}

In Sections ~\ref{sec:lc}-\ref{sec:gb}, we start by analyzing two archetypal ellipsoidal systems: particles in different liquid crystalline (LC) phases, and particles governed by the classical Gay-Berne potential. These case studies demonstrate the similarities and differences between \anisoap and traditional LC order parameters, namely the orientational (or nematic) order parameter and pair-wise Steinhardt order parameters, and examine the ability of \anisoap to perform continuous supervised tasks.
In Section~\ref{sec:benzene}, we then move to the opposite end of the molecular spectrum and analyze benzene crystals -- molecules that are ellipsoidal in shape but whose atomic interactions are complex and fundamental to their interaction landscape. Of the systems that ellipsoidal \AniSOAP is already well-suited to describe, these cases provide a relative upper- and lower-bound to model performance from a system where interactions are, by definition, entirely defined by the ellipsoidal correlations to those where the energetics are minimally determined by the shape of the molecule alone.

Unless otherwise specified, when discussing $X_\anisoap$, we are referring to the power spectrum (the \textit{3-body} term) given in 
Eq.~\ref{eq:3b}, but note the simplicity of continuing onto higher body-order terms for greater accuracy and transferability\cite{nigam_recursive_2020}.

Liquid crystals (LC) are mesoscopic phases that form from rigid ellipsoid-like particles or molecules and can be controlled to direct the flow of light for a variety of applications.\cite{hamley_introduction_2007} 
Typically, LCs are characterized as phases with orientational order but limited positional or translational order. Molecules with orientational order but no translational order are generally considered ``nematic'', whereas those exhibiting one direction of translational order (e.g. the particles appear as stacking planes, where within the planes, there is little translational order) are deemed \textit{smectic}.

In this analysis, we aim to show how our new AniSOAP representation is similar to previously established descriptors and how it enables functionality and analyses beyond the current capabilities.
In characterizing LC phases, scientists have relied on a library of different order parameters, including different orientational order parameters \cite{ramasubramani_freud_2020} to classify the orientational alignment of particles,  Steinhardt order parameters (SOPs \cite{steinhardt_bond-orientational_1983}) to characterize the neighborhood of different particles. We note the similarity of SOPs to Eq.~\eqref{eq:dens_exp}, wherein an SOP for a given particle involves the integration of spherical harmonics over that particle's neighbors:

\begin{equation}
\braket{lm|q_i}=\frac{1}{N_j}\sum_{j} w_{ij} \ylm{r_{ij}}
\end{equation}

where $j$ is a neighbor of $i$, and weights $w_{ij}$ can be introduced based on neighbor distances or Voronoi tesselations\cite{lechner_accurate_2008}. The $l^{th}$ SOP is computed by combining these terms such that 

\begin{equation}
X_{SOP} = \braket{l|q_i} = \sqrt{\frac{4\pi}{2l+1} \sum_{m=-l}^{l} |\braket{lm|q_i}|^2 }.
\end{equation}
Thus, for comparison, for all configurations, we also compute the orientational order parameter $\bX_{OOP}$ (traditionally called the nematic order parameter) ~and distance-weighted Steinhardt order parameters $\bX_{SOP}$ with  \texttt{freud}\cite{ramasubramani_freud_2020}, using similar $l_\text{max}$ and cutoff radius to ensure comparability. All representations are properly scaled and centered -- that is, each matrix of feature vectors is centered to have zero mean with unit variance, where $\bX_{SOP}$ and $\bX_\anisoap$ are scaled non-column-wise in order to retain important relative variance information.

We generate 1000 liquid crystal configurations, 500 exhibiting nematic order, and 500 exhibiting smectic order. Phases were generated to correspond with a range of orientational order parameters, from 0 (particles are randomly oriented) to 1 (particles are all ordered along the same director). We populate each of these configurations with prolate ellipsoids with a length-to-diameter ratio (L/D) equal to 3. For each of these configurations, we compute the \anisoap radial and power spectrum $\sigma_{GTO}=3.0$, $n_\text{max}=6$, $l_\text{max}=6$ (for the power spectrum), and a cutoff sufficient to include the surrounding neighbors of each ellipsoid, including in neighboring smectic planes. 

% While the suite of these order parameters (and their higher-order correlations) are a powerful set of tools for crystal structure identification and classification\cite{spellings_machine_2018}, they have yet to be fully leveraged for other supervised tasks. 
We first look at the principal components (PCs) of the \AniSOAP vectors for these phases (Fig.~\ref{fig:lc_pca}). The PCA reflects that the \anisoap features can be used to delineate translational order (blue versus red coloring), and orientational order (saturated versus unsaturated color). 
These mappings highlight qualitatively the information contained in the \anisoap representation and show that it simultaneously and smoothly represents the translational and orientational order of the configurations. In Fig.~\ref{si:lc_pca_hetero}, we demonstrate that the \anisoap features also smoothly delineate different particle shapes, with a similar plot provided with data for both L/D=2 and 3 ellipsoids. By combining this information into one smooth feature space, we are able to better recreate the ``nearsightedness'' of molecular interactions\cite{prodan_nearsightedness_2005}, as these aspects (shape, mutual orientation, and translational order) are often interrelated and correlated in the ways in which they influence molecular behavior. 

\begin{figure}[h]
    \centering
    \includegraphics[width=\linewidth]{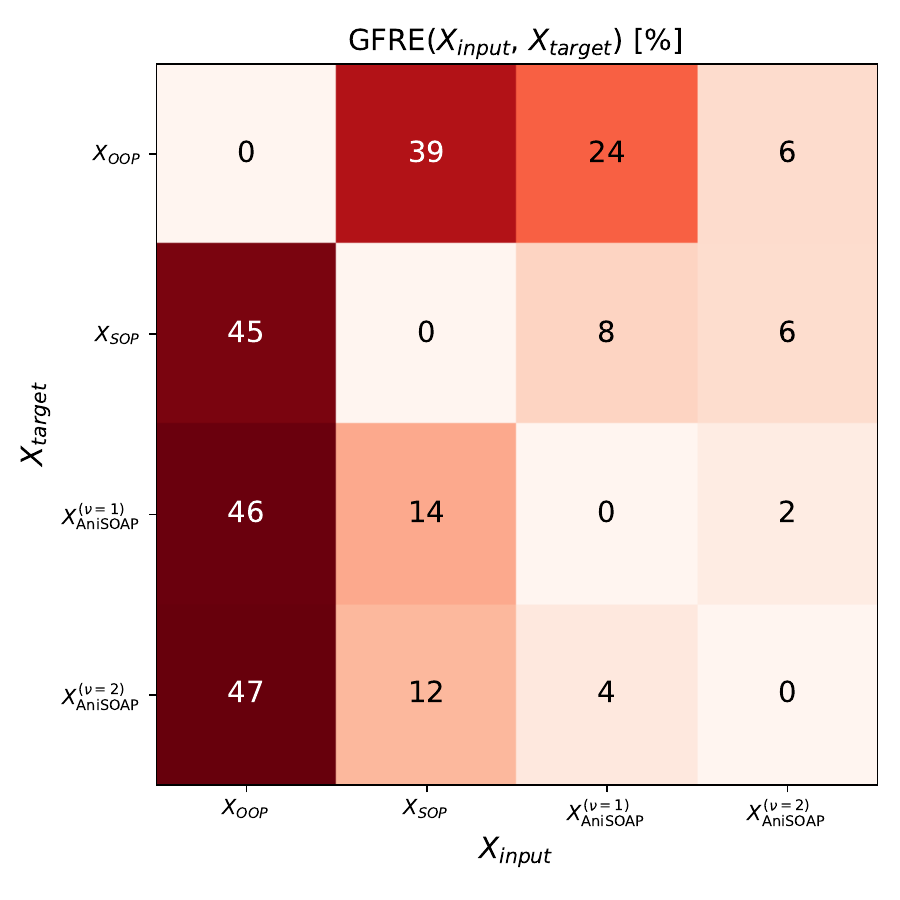}
    \caption{\textbf{Comparison of Different Featurizations for Liquid Crystal Configurations.} The Global Feature Reconstruction Error (GFRE) encodes the error using one feature representation $\bX_\text{input}$ to reconstruct another $\bX_\text{target}$, and can be used to infer the information density of one featurization compared to another. Higher values indicate poor reconstruction, whereas lower indicates better reconstruction. Here we see that \anisoap features better reconstruct traditional Steinhardt order parameters ($q_{0-6}$) and orientational order parameters than vice versa. While Steinhardt OPs carry large mutual information with the \anisoap vectors, they are unable to reconstruct the Orientational Order Parameter.}
    \label{fig:lc_gfre}
\end{figure}

Quantitatively, we can compare the information density of \anisoap with other order parameters for our configurations using the global feature reconstruction error (GFRE\cite{goscinski_role_2021}, computed using \texttt{scikit-matter}\cite{goscinski_scikit-matter_2023}). For two representations $\bX_1$ and $\bX_2$ of the same dataset, $\text{GFRE}(X_1, X_2)$, determines how much information $\bX_1$ contains relative to $\bX_2$, where 0.0 infers $\bX_1$ can perfectly reconstruct $\bX_2$, and 1.0 corresponds to poor reconstruction. As shown in Fig.~\ref{fig:lc_gfre}, we see that \anisoap representations (both the 2-body radial spectrum $X^{(\nu=1)}_\anisoap$ and 3-body power spectrum $X^{(\nu=2)}_\anisoap$) are able to reconstruct the analogous $\bX_{SOP}$ with lower error than vice versa (6-8\% versus 12-14\% reconstruction error). Furthermore, the \anisoap representations are able to reconstruct the orientational order parameter with low error (6\% for the 3-body, and 24\% for the 2-body, compared to 39\% for $X_{SOP}$). The improvement in information density can be attributed to the smooth radial bases that underlie the \anisoap construction. It is worth noting that the Steinhardt order parameter is not well-suited for heterogenous datasets, as shown in Fig.~\ref{si:lc_gfre}, and its ability to reconstruct the \anisoap representation will decrease when a dataset contains multiple particle types.

\subsection{Gay-Berne Ellipsoids}
\label{sec:gb}

The Gay-Berne (GB) potential is a Lennard Jones-type potential that contains additional terms to account for ellipsoidal anisotropies. We use the generalized formulation of the Gay-Berne potential outlined by \citet{everaers_interaction_2003} and first introduced by \citet{berardi_generalized_1995}. 
This formulation calculates the pairwise potential between (potentially dissimilar) ellipsoids, i.e., ellipsoids with three unequal semi-axes: $a_i, b_i, c_i$. These three semi-axes define a diagonal structure matrix for particle i:

\begin{equation}
    \bS_i = \begin{pmatrix}
        a_i &0 &0 \\ 
        0 &b_i &0 \\ 
        0 &0 &c_i \\ 
    \end{pmatrix}
\end{equation}
where $\bS$ is analogous to $\bD^{-2}$ of Eq.~\eqref{eq:gauss_D}, although $(a,b,c)$ are  discrete semiaxes, rather than Gaussian widths, like $\sigma$. In our case-study, we generate 25,000 dimers of $a_i=1, b_i=1.5, c_i=2$ ellipsoids at random offsets and orientations.

The center position and orientation of ellipsoid $i$ is given by position vector $\br_i$ and a 3$\times$3 rotation matrix $\rot_i$, respectively. The relative position is $\br_{12} = \br_2 - \br_1$. The Gay-Berne potential takes into account interparticle distance dependence and orientation dependence of the ellipsoid pairs. The potential is given as a product of three terms: 

\begin{align}
    % U(\boldsymbol{A_1}, \boldsymbol{A_2}, \Vec{r_{12}}) = U_r(\boldsymbol{A_1}, \boldsymbol{A_2}, \Vec{r_{12}}) * 
    U(\rot_1, \bS_1, \rot_2, \bS_2, \vec{r_{12}}) = \underbrace{U_r(\rot_1, \bS_1, \rot_2, \bS_2, \vec{r_{12}})}_\text{LJ-like term}\\\cdot\underbrace{\eta_{12}(\rot_1, \bS_1, \rot_2, \bS_2)\cdot\chi_{12}(\rot_1, \rot_2, \hat{r_{12}})}_\text{anisotropy corrections}
\label{eq:gb}
\end{align}
The first term $U_r$ is given as follows:

\begin{equation}
    U_r = 4\epsilon_{GB}\left[\left(\frac{\sigma_{GB}}{h_{12}+\gamma\sigma_{GB}} \right)^{12} - \left(\frac{\sigma_{GB}}{h_{12}+\gamma\sigma_{GB}} \right)^{6}\right]
\end{equation}
where the distance $h_{12}$ between two ellipsoids is estimated as the Perram distance of closest approach\cite{perram_statistical_1985}
\begin{equation}
    h_{12}(\rot_1, \rot_2, \bS_1, \bS_2, \br_{12}) = r_{12} - s_{12}(\rot_1, \rot_2, \bS_1, \bS_2, \hat{r}_{12})
\end{equation}
where
\begin{equation}
    s_{12}(\rot_1, \rot_2, \bS_1, \bS_2, \hat{r}_{12}) = \left[\frac{1}{2}\hat{r}_{12}^T \boldsymbol{\Xi}_{12}^{-1}(\rot_1, \rot_2,\bS_1, \bS_2) \hat{r}_{12}\right]^{-1/2}
\end{equation}
and
\begin{equation}
    \boldsymbol{\Xi}_{12}(\rot_1, \rot_2, \bS_1, \bS_2) = \rot_1^T\bS_1^2\rot_1 + \rot_2^T\bS_2^2\rot_2
\end{equation}

The terms $\epsilon_{GB}$ and $\sigma_{GB}$ are the potential well depth and location analogous to LJ. $\gamma$ is a parameter that shifts the potential well and is typically set to 1. Since the interparticle distance is defined to be the ``distance of closest approach'', $h_{12}$, rather than center-center distance, so $U_r$ must also contain dependence on the geometry and orientation of each ellipsoid to calculate $h_{12}$. 
The remaining pieces of Eq.~\eqref{eq:gb} account for the ellipsoidal geometry ($\eta_{12}$) and misorientation ($\chi_{12}$); further explanation of these terms are given in Sec.~\ref{si:gb}.

\begin{figure}
    \centering
    \includegraphics[width=\linewidth]{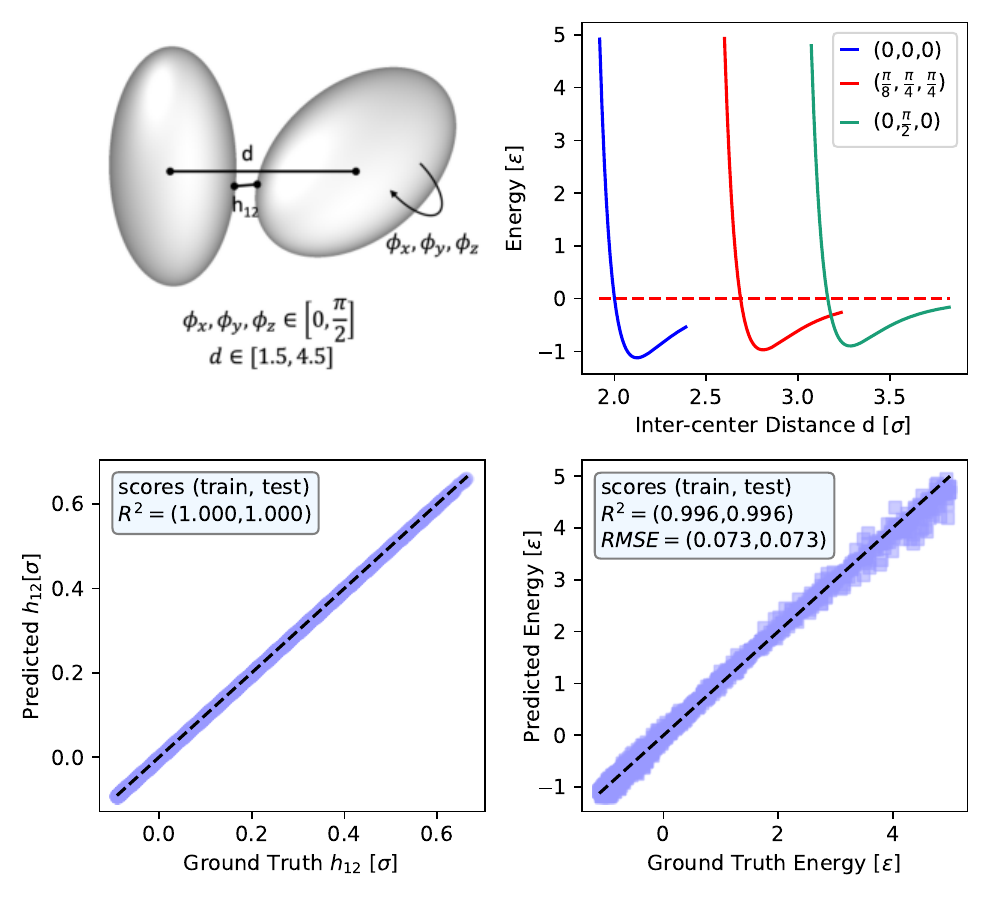}
    \caption{\textbf{Predictions of ellipsoid interactions based on the Gay-Berne Potential}. The ellipsoid dimer dataset consists of two ellipsoids with varying inter-center distance d and orientations based on rotations along the x ($\phi_x$), y ($\phi_y$), and z ($\phi_z$) axes. These varying orientations and distances lead to different characteristic energy wells, shown in the top right. Parity plots detail the performance in predicting the distance of closest approach $h_{12}$ (bottom-left, units of characteristic length scale $\sigma$) and energies (bottom-right, units of potential well depth $\epsilon$). For clarity, only the test points are shown; the train points show very similar trends.}
    \label{fig:gb_learning}
\end{figure}

One of the less trivial aspects of the Gay-Berne implementation is calculating the ``distance of closest approach'' $h_{12}$. The approximation given by \citet{perram_statistical_1985} is computationally efficient, but open to failure in some cases, like in the case of two bodies with very unequal radii \cite{everaers_interaction_2003}. It is worth noting that it is the nature of density expansions to implicitly contain the distance of closest approach, provided that the chosen cutoff distance and radial resolution are sufficiently large to include any necessary neighbors, which is evidenced by the perfect agreement in the computed and predicted $h_{12}$ distances given in Fig.~\ref{fig:gb_learning}\footnote{The $h_{12}$ distances here fall within the confidence region of the \citet{perram_statistical_1985} algorithm, so they can be taken as exact}. And thus, in \anisoap, we alleviate the requirement to explicitly compute $h_{12}$ or rely on its approximations.

From here, learning the Gay-Berne potential is straightforward, similar to how traditional SOAP vectors can learn the LJ potential. With the \anisoap power spectrum vectors, it is possible to interpolate, using solely regularized linear regression, the interaction potential of ellipsoids of arbitrary distances from one another, as shown in Fig.~\ref{fig:gb_learning}. Performance decreases for repulsive, heavily overlapped configurations (right end of the parity plot in the lower right of Fig.~\ref{fig:gb_learning}), likely due to the fact that, in the repulsive regime, small feature differences correspond to disproportionately large energy differences. Our learning exercise converges with as few as 1,000 training points, as evidenced by the learning curve given in Fig.~\ref{si:gb_learning_curve}.

\subsection{Benzene Molecules}
\label{sec:benzene}

\begin{figure}[h]
    \centering
    \includegraphics[width=\linewidth]{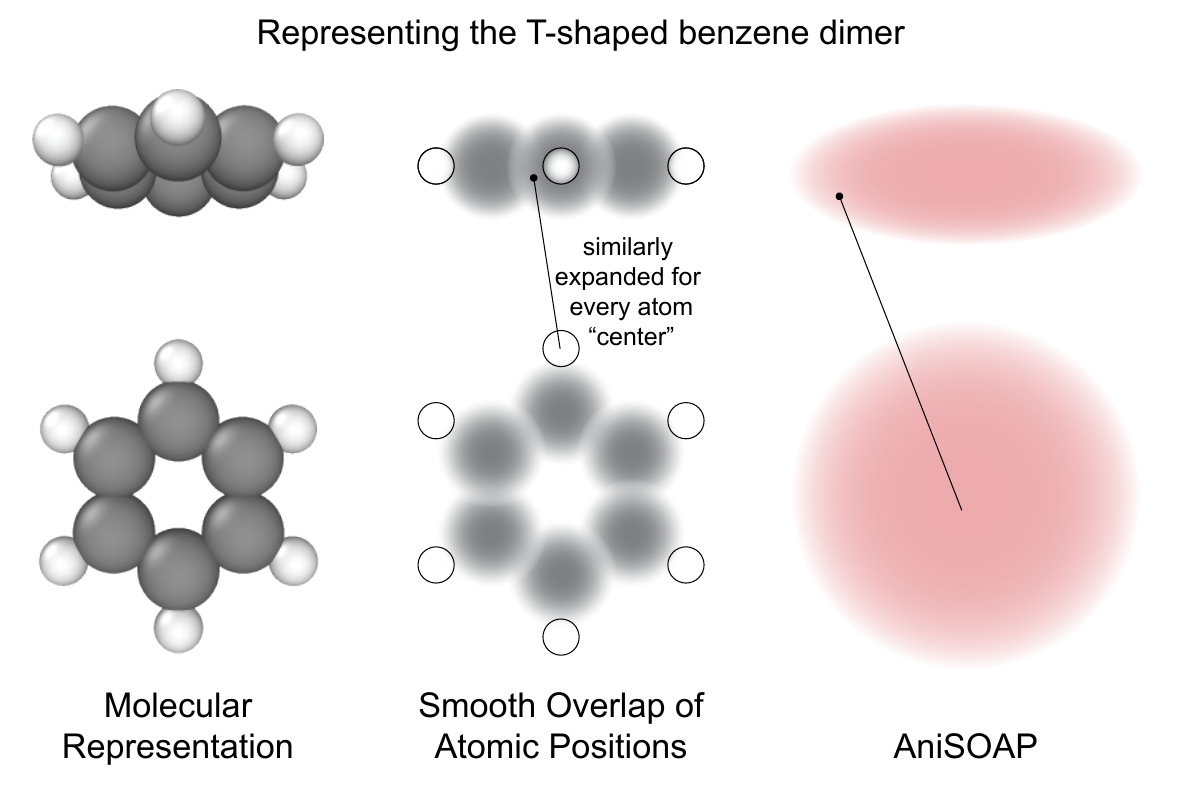}
    \caption{Representations of the T-Shaped Benzene Dimer. From left to right. A molecular image of the dimer at approximately 5.5\AA~separation, generated using Ovito\cite{stukowski_visualization_2009}. The underlying representation to the SOAP formalism, wherein each atom is represented by a Gaussian field, and we expand n-body terms of each atomic neighborhood. The underlying representation for the AniSOAP formalism, wherein each \emph{molecule} is represented by an anisotropic Gaussian field, and we expand n-body terms of each \emph{molecular} neighborhood. }
\end{figure}

For the following analysis, we intentionally construct a set of benzene configurations to demonstrate the possibilities and, simultaneously, the limitations for \anisoap.
% we use two distinct datasets of benzene conformations:
% one containing configurations of non-periodic benzene dimers and the other randomly generated periodic benzene crystals. These datasets are meant to provide both an interpretable and practical look at the use of \anisoap.

% The dimers are taken from the results published \citet{sinnokrot_highly_2004} that report energetics determined by second-order Moller-Plesset perturbation theory and coupled-cluster with singles, doubles, and perturbative triples with a modified augmented, correlation-consistent basis set (the so-called ``CCSD(T)/aug-cc-pVQZ*'' level of theory). These results, developed to guide the study of weakly-bound $\pi-\pi$ systems typically poorly described by traditional electronic structure theory (with errors on the order of 2.0 kcal/mol)\cite{sinnokrot_highly_2004}, contain 118 benzene dimer configurations in the sandwich, T-shaped, and parallel-displaced configurations and their binding energies. This dataset is too small to sufficiently demonstrate model efficacy, so we rely on it solely for demonstrative purposes. 
We start with a set of symmetry-constrained crystals comprised of planar benzene molecules with the software \texttt{PyXtal}\cite{fredericks_pyxtal_2021} across all 230 space groups. We then augmented this dataset by randomly rotating or translating the molecules, where many configurations correspond to the same \emph{positions} or \emph{orientations} of the molecules. 
The resultant dataset contains roughly 7,000 benzene crystals. We then compute energetic quantities using QuantumEspresso v7.0 \cite{giannozzi_quantum_2009} using Perdew–Burke-Ernzerhof (PBE) pseudopotentials and cutoff parameters reported by \citet{prandini_precision_2018}, a Grimme D3-dispersion correction\cite{grimme_dispersion-corrected_2016}, and a 3x3 Monkhorst-Pack k-point grid\cite{monkhorst_special_1976}. Computations were managed with the \texttt{signac} and \texttt{signac-flow} packages\cite{ramasubramani_signac_2018, adorf_simple_2018}, and computed using the \texttt{ASE} computational front-end\cite{hjorth_larsen_atomic_2017}. %For reproducibility, example scripts are included in the supplementary information.

\subsubsection*{Hyperparameter Tuning} 

With any new featurization, there is always valid concern about hyperparameter optimization and the sensitivity of the featurization to changes in these values\cite{cheng_mapping_2020}. In many ways, this is the appeal of the SOAP formalism -- the hyperparameters, namely the widths of the Gaussian densities, basis sets, and cutoff parameters, all have roots in chemical physics and can be chosen from a large range of ``reasonable'' values with minimal impact on model interpretability and performance.

As the primary axes of planar benzenes are well-defined, the only new hyperparameters to consider are the semiaxes lengths of the MVG. For simplicity, we will only consider $\sigma_1=\sigma_2$, and prove that, in line with conventional knowledge on benzene geometry, best results are obtained with $\sigma_1>\sigma_3$ (an oblate ellipsoid). 

To tune these parameters, we again use the GFRE of a given \anisoap representation and an analogous SOAP representation of our benzene crystal configurations. By taking $\text{GFRE}(X_\text{AniSOAP}, X_\text{SOAP})$, we determine how much information is lost by moving from the atomistic to coarse-grained \anisoap representation, again where 0.0 infers perfect reconstruction of the atomic correlations, and higher numbers correspond to poor reconstruction. Note that this hyperparameter tuning occurs independently of any traditional supervised learning task.

For our SOAP representation, we compute the structure-averaged representation using \texttt{rascaline}\cite{fraux_rascaline_2023}, with a cutoff radius of $7.0$\r{A}, Gaussian density width of $0.5$\r{A}, $l_\text{max}=10$, and $n_\text{max}=4$. This results in a length $1,188$ vector for each configuration. We compare against \anisoap vectors using a similar cutoff radius, number of angular and radial channels, varying $\sigma_1=\sigma_2$, $\sigma_3$, and $\sigma_\text{GTO}$. Each \anisoap vector has length $146$.

\begin{figure}
    \centering
    \includegraphics[width=\linewidth]{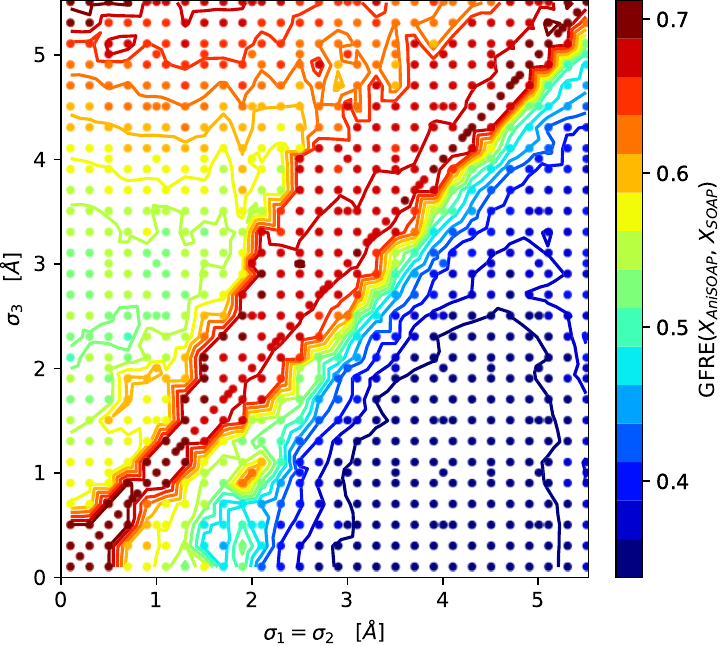}
    \caption{\textbf{Results of tuning $\sigma_1$ and $\sigma_3$ for benzene crystals.} The color of the scatter points and contour levels denote $\text{GFRE}(X_\text{AniSOAP}, X_\text{SOAP})$, where lower values indicate a greater fidelity of the \anisoap representation to the SOAP analog, and higher values indicate a greater information loss. Our MVG is oblate where $\sigma_1>\sigma_3$, prolate where $\sigma_1<\sigma_3$, and spherical where $\sigma_1=\sigma_3$. As expected, oblate representations carry more mutual information to the atomistic representation than prolate or spherical ones. The spherical representation $\sigma_1=\sigma_2=\sigma_3$ is consistent with constructing a traditional SOAP representation from the molecule centers.}
    \label{fig:gfre}
\end{figure}

From the results in Fig.~\ref{fig:gfre}, we see how the oblate MVG with $\sigma_1=\sigma_2=$4.0 and $\sigma_3=$0.5 minimizes the information loss compared to the atomistic SOAP representation with a $\sigma_{GTO}=1.5$. Prolate ($\sigma_1<\sigma_3$) and spherical ($\sigma_1=\sigma_3$) Gaussian densities perform notably worse. It is interesting, however, that there is a wide range of $\sigma_1, \sigma_3$ values that obtain similar results, signifying that the \anisoap hyperparameters are robust with respect to small changes, provided the MVG remain oblate, and the semiaxis lengths are within reasonable proportion. We include in the SI similar analyses for $\sigma_\text{GTO}=\{0.5, 1.0, 1.5, 2.0, 2.5, 3.0\}$, demonstrating that $\sigma_\text{GTO} \in [1.0\text{\r{A}}, 3.0\text{\r{A}}]$ obtains similar results to $\sigma_\text{GTO}=1.5$\r{A}.

\subsubsection*{Learning of Benzene Energetics} 
Taking the optimized hyperparameters from the previous section, we now show how the \anisoap representation performs in simple supervised tasks, demonstrating where \anisoap performs best and where it is incomplete for atomistic systems. It is worth noting that the intention of \anisoap is to provide a molecule-level analog to SOAP, and is, by design, incomplete\cite{pozdnyakov_completeness_2020} with respect to many atom-atom correlations.

\begin{figure*}
    \centering
    \includegraphics[width=\linewidth]{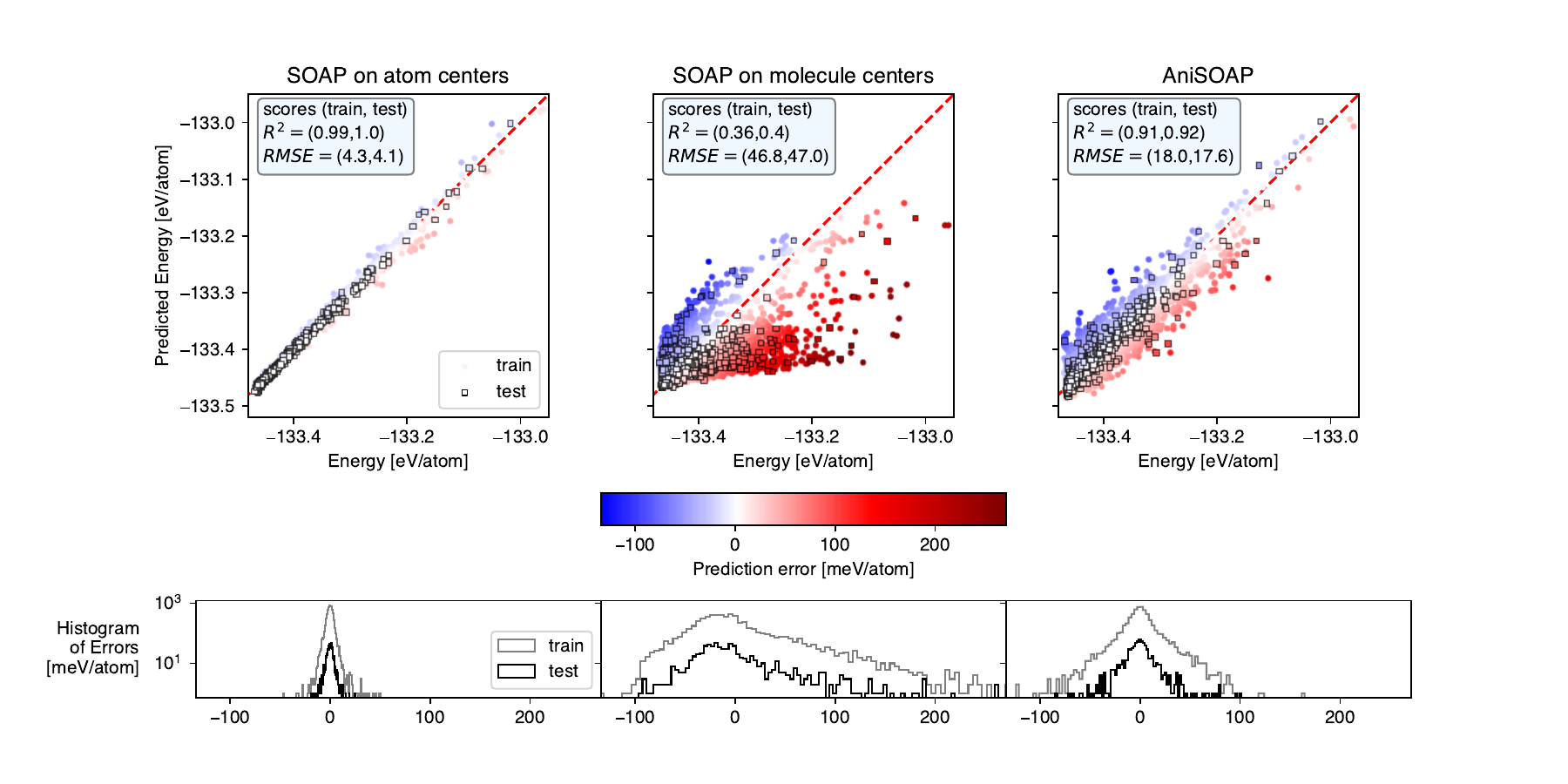}
    \caption{\textbf{Parity plots comparing the performance of SOAP (built from the atom centers), optimized SOAP (built on the molecule centers), and \anisoap} for the dataset of benzene crystalline configurations. All regressions were conducted using 5-fold cross-validated regularized linear regression, and errors are reported for the same 90/10 training/test split. The SOAP representations was first reduced via PCA to be of similar rank to \anisoap. In the parity plots, color denotes the error in meV/atom, with the corresponding distribution of errors shown in the lower panel. 
    }
    \label{fig:parity}
\end{figure*}

To highlight the importance of molecular representation, we will use regularized ridge models, employing a 90/10 training/test split and five-fold cross-validation. We first perform three regression tasks -- learning the baselined, per-atom energy (equivalent to learning the per-molecule energy) using the all-atom SOAP representation, a  SOAP representation built solely from the molecule centers (coinciding with the ridge line at $\sigma_1=\sigma_2=\sigma_3$ in Fig.~\ref{fig:gfre}), and our optimized \anisoap representation. We have enforced that all three representations have similar ranks, as higher-rank features will outperform lower-rank ones based on size alone. To do so, we perform dimensionality reduction via principal components analysis\footnote{principal covariates regression\cite{helfrecht_structure-property_2020}, while ideal for regression-related dimensionality reductions, would again limit our ability to compare the \emph{relevant information content} of these representations.}.

The results, as shown in Fig.~\ref{fig:parity}, demonstrate how incorporating the anisotropy of intermolecular correlations can greatly improve our ability to learn energetic quantities, even in contexts where atom-atom interactions are important. A SOAP representation on the molecule centers (center panels) \footnote{using the same major semiaxis length as the width of the isotropic Gaussian field, and keeping all other hyperparameters consistent}, has heavily limited regression performance, with a typical RMSE on the order of $45-50$meV/atom, and an $R^2\leq0.4$. By introducing intermolecular anisotropy, our performance jumps $R^2\approx0.9, \text{RMSE}\approx 15-20$meV/atom, much closer to the all-atom regression in the left panel of Fig.~\ref{fig:parity}.

\begin{figure}
    \centering
    \includegraphics[width=\linewidth]{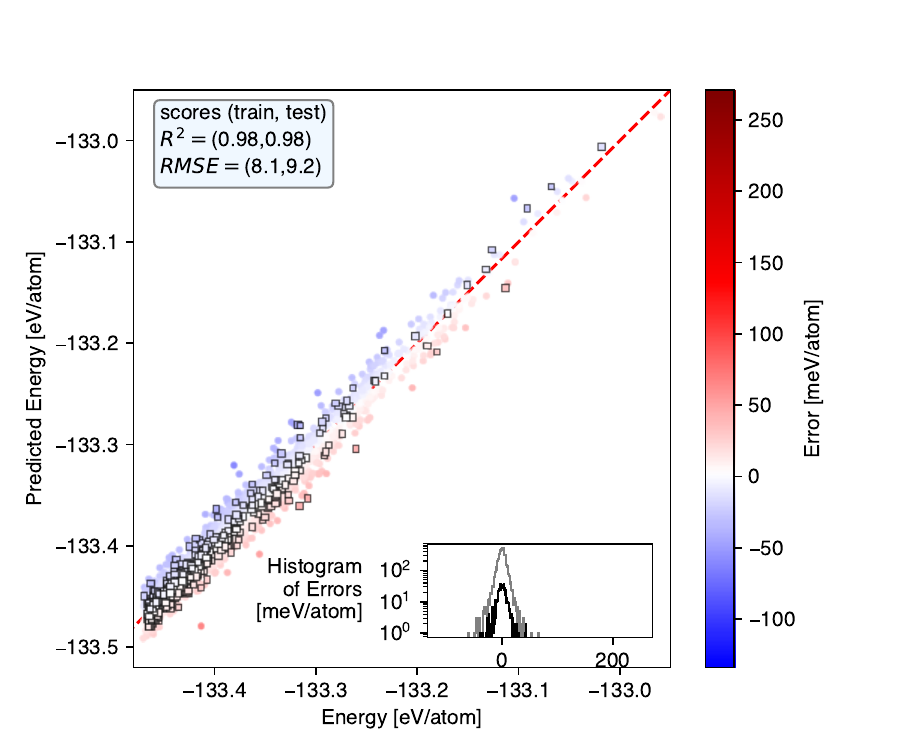}
    \caption{\textbf{A parity plot showing the results of baselining our energies on the \anisoap approximation, followed by atom-level regression via SOAP.} The inset demonstrates the improvement in the distribution of errors.
    }
    \label{fig:correction}
\end{figure}

Given that, by definition, \anisoap does not contain the information on atomistic correlations, it follows naturally that we can learn a ``correction'' to the \anisoap energies from our original SOAP representation, as corroborated by Fig.~\ref{fig:correction}. This is important for a few reasons -- firstly, it shows that \anisoap can be used as a coarse-graining baseline to complex energetics. Simply learn the energies due to body-body interactions, and incorporate an atom-level correction where necessary. Secondly, it helps identify these spaces where atom-atom interactions are driving the majority of energetic quantities, which can be confirmed by leveraging the linear relationship between the atom-level SOAP vector representation and the errors in the \anisoap regression and perform Principal Covariates Regression (PCovR)\cite{helfrecht_structure-property_2020, goscinski_scikit-matter_2023}. We have included in the SI such a mapping, explorable via the online tool \href{chemiscope.org}{chemiscope}\cite{fraux_chemiscope_2020}. We also posit that one can do a combined \anisoap and SOAP representation of a system to optimize the segmentation of molecular energetics; however, this is beyond the scope of this current study.

\section{Conclusions}
The tools developed under the umbrella of ``machine-learning'' are powerful, and revolutionizing society and the scientific community at an aggressive pace. One important advancement for chemical sciences is the concept of treating chemistry as data -- choosing specifically the numerical representations with which to cast chemical questions into statistical and analytical models. These representations are strongest when grounded in the established physical principles that we are trying to emulate or predict.

Here, we have demonstrated one such physically-driven approach to machine-learning representations, aimed at leveraging machine learning for coarse-grained and mesoscale entities, demonstrating such an approach for particles and molecules that are well-represented by ellipsoidal bodies. Our results show that, for both classical and quantum mechanical datasets, \anisoap provides a suitable representation for ellipsoidal bodies, as these representations are able to accurately and \emph{linearly} map onto complex energetics without the need for deeper ML infrastructures. We have constructed \anisoap in such a way as to retain compatibility with the popular atomistic SOAP formalism -- in practice, we hope that these two technologies are used hand-in-hand to simplify energetic landscapes and explicitly incorporate many-body effects.
Furthermore, this consistency between formalisms can enable multiscale machine learning approaches, wherein one can use coarse-grained representations for long-scale molecular motion and atom-atom representations for nearsighted interactions. Future work will focus on the efficient implementation of anisotropic decompositions for many-body effects and their application to such systems.

\section{Data availability}
Data for this paper will be made available via MaterialsCloud\cite{talirz_materials_2020} upon acceptance, including all raw input files and analysis scripts.

\section{Author contributions}
KKHD derived the expansion of multi-variate Gaussian and wrote the corresponding proof-of-principle code.
     AL, JN, Y-CC and RKC refined the code, implementing orthonormalization and additional basis sets.
AL and RKC wrote the manuscript and designed and executed the case studies.

\section{Conflicts of interest}
There are no conflicts to declare.

\section{Acknowledgements}
This project was funded by the Wisconsin Alumni Research Fund (RKC), by NSF through the University of Wisconsin Materials Research Science and Engineering Center (DMR-2309000, AL), and the European Research Council (ERC) under the research and innovation program (Grant Agreement No. 101001890-FIAMMA, JN, KKHD).

We extend our un-ending gratitude to Guillaume Fraux and the developers of \href{https://github.com/Luthaf/rascaline}{\texttt{rascaline}} for fielding our many questions during the implementation and validation of \anisoap~.
\bibliography{references}

\appendix

\renewcommand\thefigure{S\arabic{figure}}
\setcounter{figure}{0}

\onecolumngrid

\section{Evaluating the Coefficients}
\label{si:evaluating}

We have shown that, by choosing a MVG density and suitable basis functions, we can analytically compute the coefficients 
\begin{align}
    \langle nlm | \rho_i \rangle,
\end{align}
by evaluating of integrals of the form
\begin{align}
    I_{nlm} = \int_{\relc}\mathrm{d}^3r e^{-\frac{1}{2}(\br-\ba)^TA(\br-\ba)} r^{2n}R_l^m(\br).
\end{align}

In this section, we will present an explicit algorithm to obtain all $I_{nlm}$. The algorithm will require two key ingredients, namely converting the spherical formulation of $r^{2n}R_l^m(\br)$ into Cartesian coordinates (\ref{sec:sphcart}), then computing the corresponding moments across the subject to our MVG (\ref{sec:moments}).

\subsection{Computing the effective quadratic form in the exponential}
\noindent If we use the monomial basis, the integrand in 
\begin{align}
    I_{nlm} & = \int_{\relc}\mathrm{d}^3r e^{-\frac{1}{2}(\br-\br_{ij})^TA(\br-\br_{ij})} r^{2n}R_l^m(\br)
\end{align}
is already in a convenient form, since the Gaussian part is completely specified by the center $\br_{ij}$ and the precision matrix $A$. If, on the other hand, we use the GTO basis, we get an extra exponential factor
\begin{align}
    I_{nlm} & = \int_{\relc}\mathrm{d}^3r e^{-\frac{1}{2}(\br-\br_{ij})^TA(\br-\br_{ij})} r^{2n}R_l^m(\br)e^{-\frac{\br^2}{2\sigma^2}} \\
    & = \int_{\relc}\mathrm{d}^3r e^{-\frac{1}{2}(\br-\br_{ij})^TA(\br-\br_{ij}) -\frac{\br^2}{2\sigma^2}} r^{2n}R_l^m(\br)
\end{align}
meaning that the Gaussian part
\begin{align}
    e^{-\frac{1}{2}(\br-\br_{ij})^TA(\br-\br_{ij}) -\frac{\br^2}{2\sigma^2}}
\end{align}
as a whole no longer has the convenient form. By completing the square, ignoring the global factor of $-\frac{1}{2}$ in the exponent, we can rewrite
\begin{align}
    (\br-\br_{ij})^TA(\br-\br_{ij}) + \frac{1}{\sigma^2}\br^2 & = (\br-\br_0)^T\tilde{A}(\br-\br_0)  +  c,
\end{align}
with
\begin{align}
    \tilde{A} & = A + \frac{1}{\sigma^2} \\
    \br_0 & = \tilde{A}^{-1}A\br_{ij} = \br_{ij} - \frac{1}{\sigma^2}\tilde{A}^{-1}\br_{ij} \\
    c & = \frac{1}{\sigma^2} \br_{ij}^T\tilde{A}^{-1}A\br_{ij}
\end{align}
The second form of $\br_0$ is more convenient to obtain a qualitative picture: the second term represents the deviation of the center due to the addition of the second Gaussian. Depending on how sharp this Gaussian is, the relative importance of this term will change.\\

Firstly, note that using the first representation of $\br_0$ and the fact that both $A$ and $\tilde{A}$ are symmetric, we get

\begin{align}
    \br_0^T\tilde{A}\br_0 & = \br_{ij}^TA\tilde{A}^{-1}\tilde{A}\tilde{A}^{-1}A \br_{ij} \\
    & = \br_{ij}^TA\tilde{A}^{-1}A \br_{ij} \\
    & = \br_{ij}^TA\tilde{A}^{-1} \left(\tilde{A} - \frac{1}{\sigma^2} \right) \br_{ij} \\
    & = \br_{ij}^TA \br_{ij} - \frac{1}{\sigma^2} \br_{ij}^T\tilde{A}^{-1}A\br_{ij}
\end{align}

Expanding the quadratic terms, we get
\begin{align}
    (\br-\br_0)^TA(\br-\br_0) & = \br^T \tilde{A} \br - 2 \br^T \tilde{A}\br_0 + \br_0^T\tilde{A}\br_0 \\
    & = \br^T A \br + \frac{1}{\sigma^2}\br^2 - 2\br^T \tilde{A}\left(\br_{ij} - \frac{1}{\sigma^2}\tilde{A}^{-1}\br_{ij}\right) + \br_0^T\tilde{A}\br_0 \\
    & = \br^T A \br + \frac{1}{\sigma^2}\br^2 - 2\br^T \tilde{A}\br_{ij} + 2\frac{1}{\sigma^2}\br^T \br_{ij} + \br_0^T\tilde{A}\br_0 \\
    & = \br^T A \br + \frac{1}{\sigma^2}\br^2 - 2\br^T A\br_{ij} + \br_0^T\tilde{A}\br_0\\
    & = \br^T A \br + \frac{1}{\sigma^2}\br^2 - 2\br^T A\br_{ij} + \br_{ij}^TA \br_{ij} - \frac{1}{\sigma^2} \br_{ij}^T\tilde{A}^{-1}A\br_{ij} \\
    & = (\br - \br_{ij})^T A (\br - \br_{ij}) + \frac{1}{\sigma^2}\br^2 - \frac{1}{\sigma^2} \br_{ij}^T\tilde{A}^{-1}A\br_{ij}
\end{align}
Thus, we do see that indeed,
\begin{align}
    (\br-\br_{ij})^TA(\br-\br_{ij}) + \frac{1}{\sigma^2}\br^2 & = (\br-\br_0)^T\tilde{A}(\br-\br_0) +  \frac{1}{\sigma^2} \br_{ij}^T\tilde{A}^{-1}A\br_{ij}.
\end{align}
% In the current implementation, the change in magnitude due to the constant $c=\frac{1}{\sigma^2} \br_{ij}^T\tilde{A}^{-1}A\br_{ij}$ is assumed to be subtracted via feature standardization.
\subsection{Spherical to Cartesian Transformation}
\label{sec:sphcart}
Recall that for both the monomial and GTO bases, our integrand is a product of a Gaussian and an expression of the form
\begin{align}
    r^{2n}R_l^m(\br) = \mathrm{poly}(x,y,z),
\end{align}
where we emphasize that $r^{2n}R_l^m(\br)$ is simply a polynomial in the variables $(x,y,z)$.\\
For practical evaluations, however, it does not suffice to know that it is equal to ``some'' polynomial. We will need to explicitly express the solid harmonics $R_l^m(\br)$ in monomial terms. 
% For instance, using the real solid harmonics, some concrete examples are
% \begin{align}
%     R_{2,0}(\br) & = \frac{1}{2} (3z^2-r^2) \\
%     & = \frac{1}{2}(2z^2-x^2-y^2) \\
%     & = z^2 - \frac{1}{2}x^2 - \frac{1}{2}y^2 \\
%     R_{2,2}(\br) & = \frac{\sqrt{3}}{2}(x^2-y^2) \\ 
%     R_{3,2}(\br) & = \frac{\sqrt{15}}{2} (x^2-y^2)z \\
%     & = \frac{\sqrt{15}}{2} (x^2z-y^2z).
% \end{align}
In general, there will exist a decomposition
\begin{align}
    R_{l}^m(\br) = \sum_{n_0+n_1+n_2=l} T^{lm}_{n_0,n_1n_2}x^{n_0}y^{n_1}z^{n_2},
\end{align}
where $T^{lm}_{n_0,n_1n_2}$ are some coefficients. These coefficients depend only on $l,m$ and can thus be cached. For the full basis function, there will then exist coefficients $T^{nlm}_{n_0,n_1n_2}$ such that
\begin{align}
    r^{2n}R_l^m(\br) = \sum_{n_0,n_1,n_2} T^{nlm}_{n_0,n_1n_2} x^{n_0}y^{n_1} z^{n_2}.
\end{align}
 As soon as we are provided with a complete list of the $(l,n)$ pair we need, we can precompute these coefficients.

\subsection{Evaluation of Moments}
\label{sec:moments}
Once we have decomposed $r^{2n}R_l^m(\br)$ into a sum of monomial terms, we have reduced the evaluation problem to integrals of the form
\begin{align}
    \langle x^{n_0}y^{n_1}z^{n_2} \rangle = \int_{\relc}\mathrm{d}^3r e^{-\frac{1}{2}(\br-\ba)^TA(\br-\ba)} x^{n_0}y^{n_1}z^{n_2}.
\end{align}
We shall call this expression the $(n_0,n_1,n_2)$-th moment of the Gaussian, or just moment for short.

To provide some context for the terminology, ``moment'' is a term used in probability theory. If we are given some probability density (in three variables) $p(\br) = p(x,y,z)$, its $(n_0,n_1,n_2)$-th moment is defined as
\begin{align}
    \int_{\relc}\mathrm{d}^3rp(\br) x^{n_0}y^{n_1}z^{n_2}.
\end{align}
The most common notation for this in the mathematical literature is to write it as $\mathbb{E}[x^{n_0}y^{n_1}z^{n_2}]$. In the quantum field theory / statistical mechanics literature, on the other hand, the notation $\langle x^{n_0}y^{n_1}z^{n_2} \rangle$ is used more commonly instead.

Connecting this back to our integrals, we can see that the Gaussian part is almost a probability density, apart from the normalization factor, motivating the use of an analogous notation.

The normalization factor is a global factor, meaning that techniques that have been developed to evaluate moments in probability theory can be applied to our problem as well. This will allow us to compute all the moments for a given Gaussian function specified by the precision matrix $A$ and the center $\ba$.

\subsection{Putting Everything Together}
Combining the two ingredients, we can now formulate the complete algorithm for the evaluation of the above mentioned integrals, namely:
\begin{align}
    I_{nlm} & = \int_{\relc}\mathrm{d}^3r e^{-\frac{1}{2}(\br-\ba)^TA(\br-\ba)} r^{2n}R_l^m(\br) \\
    & = \int_{\relc}\mathrm{d}^3r e^{-\frac{1}{2}(\br-\ba)^TA(\br-\ba)} \sum_{n_0,n_1,n_2} T^{n,lm}_{n_0,n_1n_2} x^{n_0}y^{n_1} z^{n_2} \\
    & = \sum_{n_0,n_1,n_2} T^{n,lm}_{n_0,n_1n_2} \int_{\relc}\mathrm{d}^3r e^{-\frac{1}{2}(\br-\ba)^TA(\br-\ba)}  x^{n_0}y^{n_1} z^{n_2} \\
    & = \sum_{n_0,n_1,n_2} T^{n,lm}_{n_0,n_1n_2} \langle x^{n_0}y^{n_1}z^{n_2} \rangle
\end{align}
We can therefore see that the computation of our features is completely determined by the two ingredients mentioned before, namely:
\begin{enumerate}
    \item the transformation coefficients $T^{n,lm}_{n_0,n_1n_2}$ that express $r^{2n}R_l^m$ using monomials
    \item the moments $\langle x^{n_0}y^{n_1}z^{n_2} \rangle$ of the Gaussian
\end{enumerate}
In the following two subsections, we will explain how to perform these two steps, respectively.

% \subsection{Writing the Basis Functions using Monomials}
\subsection{Evaluation of Solid Harmonics, Spherical Harmonics Part}
In this subsection, we present an algorithm that will allow us to decompose the basis functions into its constituent monomial terms. For this, we start from the definition
\begin{align}
\label{eq:solid-harmonics}
	R_l^m(\br) = \sqrt{\frac{4\pi}{2l+1}}r^l Y_l^m(\hat{r}) = \sqrt{\frac{(l-m)!}{(l+m)!}} r^lP_l^m(\cos\theta)e^{im\phi}.
\end{align}
The $\phi$ dependent part will be a polynomial of degree $m$, while the associated Legendre polynomial will provide us with the remaining $l-m$ degrees. Starting with the exponential, we get
\begin{align}
	e^{im\phi} & = \left(e^{i\phi}\right)^m \\
	& = \left( \cos\phi + i\sin\phi\right)^m \\
	& = \sum_{k=0}^m \begin{pmatrix} m \\ k \end{pmatrix} i^{m-k}\cos^k\phi \sin^{m-k}\phi
\end{align}
After multiplying by $r^m\sin^m\theta$, this will result in contributions of the form
\begin{align}
	\sum_{k=0}^m \begin{pmatrix} m \\ k \end{pmatrix} i^{m-k}x^k y^{m-k}.
\end{align}
For the associated Legendre polynomial part, it is possible to use the recurrence relation\cite{noauthor_dlmf_nodate}
\begin{align}
\label{eq:legendre-recurrence}
	P_{l+1}^m(x) = \frac{2l+1}{l-m+1} xP_l^m - \frac{l+m}{l-m+1} P_{l-1}^m 
\end{align}
and 
\begin{align}
\label{eq:legendre-l}
	P_l^l(x) = (-1)^l(2l-1)!!(1-x^2)^\frac{l}{2}
\end{align}
The special case for $l=m$ can be used to initialize the recurrence relations by computing all the associated Legendre polynomials at $l=m$ and $l=-m$ with symmetry, while the recurrence relation can be used to generate all higher order ones.\\

Combining Eqs. ~\ref{eq:solid-harmonics}, \ref{eq:legendre-recurrence} and ~\ref{eq:legendre-l}, we get 
\begin{align}
    R_{l+1}^m &= \sqrt{\frac{(l+1-m)!}{(l+1+m)!}} \sqrt{\frac{(l+m)!}{(l-m)!}} \frac{2l+1}{l+1-m} z R_l^m - \\&\sqrt{\frac{(l+1-m)!}{(l+1+m)!}} \sqrt{\frac{(l-1+m)!}{(l-1-m)!}} \frac{l+m}{l+1-m} r^2 R_{l-1}^m
\end{align}

from where we can identify the prefactors used in the algorithm below
The equation be further simplified to obtain the recurrence as, 

\begin{align}
    R_{l+1}^m &= \sqrt{\frac{(l+1-m)}{(l+1+m)}} \frac{2l+1}{l+1-m} z R_l^m - \\&\sqrt{\frac{(l+1-m)!}{(l+1+m)!}} \sqrt{\frac{(l-1+m)!}{(l-1-m)!}} \frac{l+m}{l+1-m} r^2 R_{l-1}^m
\end{align}
\iffalse
ALGORITHM, INITIALIZATION
\begin{lstlisting}
T = []
for l in 0,1,2,...,maxdeg:
    # Initialize array in which to store all
    # coefficients for each l
    T_l = zeros((2l+1,nmax,maxdeg+1,maxdeg+1,maxdeg+1))

    for n0 in 0,1,...,l:
        n1 = l-n0
        T_l[m=l,n=0,n0,n1,n2=0] = prefac * l_choose_n0
        # Warning: in reality, m=l corresponds to the index 2l

        T_l[m=-l,n=0,n0,n1,n2=0] = (-1)^l*T_l[l,0,n0,n1,0]
        # Warning: m=-l corresponds to index 0

    T_l *= (-1)^l * (2l-1)!! * other_prefacs
    T.append(T_l)
\end{lstlisting}
ALGORITHM, RECURRENCE
\begin{lstlisting}
for l in 1,2,...,maxdeg:
    # Define the prefactor arrays 
    ms = np.arange(-l, l+1) # m=-l,-l+1,...,l
    a = prefac_1(ms) # length 2l+1 due to m dependence
    b = prefac_2(ms) # length 2l+1 due to m dependence
    for n0,n1,n2 in moment_indices:
        T[l][:,0,n0,n1,n2] += a * T[l-1][:,0,n0,n1,n2-1]
        T[l][:,0,n0,n1,n2] += b * T[l-2][:,0,n0-2,n1,n2]
        T[l][:,0,n0,n1,n2] += b * T[l-2][:,0,n0,n1-2,n2]
        T[l][:,0,n0,n1,n2] += b * T[l-2][:,0,n0,n1,n2-2]
\end{lstlisting}
\fi

\subsection{Evaluation of Radial Dependence}
If all transformation coefficients for $n=0$ are known, it is straight forward to compute those for higher $n\geq 2$ using 
\begin{align}
	B_{n+1,lm} = r^2 B_{nlm} = (x^2+y^2+z^2)B_{nlm} \equiv (x^2+y^2+z^2)r^{2n}R_l^m
\end{align}
Thus, if all transformation coefficients $T^{nlm}_{n_0,n_1,n_2}$ at some $n$ are known, we obtain those for higher $n$ by running the iteration:
\begin{align}
	\forall (l,m,n_0,n_1,n_2):& \mathrm{with} \quad n_0+n_1+n_2=l + 2n =  d\\
	& T^{n+1,lm}_{n_0+2,n_1,n_2} += T^{n,lm}_{n_0,n_1n_2}\\
	& T^{n+1,lm}_{n_0,n_1+2,n_2} += T^{n,lm}_{n_0,n_1n_2}\\
	& T^{n+1,lm}_{n_0,n_1,n_2+2} += T^{n,lm}_{n_0,n_1n_2}
\end{align}
After the iteration, all coefficients at the radial channel $n+1$ will have the correct values.\\\iffalse
ALGORITHM: RECURRENCE WITH RESPECT TO n
\begin{lstlisting}
for l in 0,1,2,...,maxdeg:
    for n in 1,2,...,nmax(l):
        deg = l + 2*n # degree of polynomial

        for n0,n1,n2 in moment_indices(l):
            # Use recurrence relation to update
            # Warning, if n0-2, n1-2 or n2-2 are negative
            # it might be necessary to add if statements
            # to avoid.
            # If we are lucky (TODO: think through this)
            # such coefficients might automatically be zero.
            T[l][:,n,n0,n1,n2,n3] += T[l][:,n-1,n0-2,n1,n2]
            T[l][:,n,n0,n1,n2,n3] += T[l][:,n-1,n0,n1-2,n2]
            T[l][:,n,n0,n1,n2,n3] += T[l][:,n-1,n0,n1,n2-2]
\end{lstlisting}
\fi

\subsection{Computing the Moments}
In this subsection, we explain how we can compute the moments
\begin{align}
    \langle x^{n_0}y^{n_1}z^{n_2} \rangle = \int_{\relc}\mathrm{d}^3r e^{-\frac{1}{2}(\br-\ba)^T\bA(\br-\ba)} x^{n_0}y^{n_1}z^{n_2}.
\end{align}
for a given precision matrix $A$ and center $\ba$.\\
As we will show, this step will be significantly easier if the matrix $\bA$ is diagonal, i.e. if we work in the basis of the principal axes. In our current implementation, we still use the original coordinate frame in which the matrix $\bA$ keeps its general form. Despite the slightly more complicated computation of the moments, this will reduce some computational cost associated with the rotation of the coordinate frame (summing over the Wigner matrix) later on. Nevertheless, there might still be some benefits to have an implementation that uses the diagonalization step. We will, therefore, start by discussing the simpler diagonal case, as well as extensions that would become possible for this special case. We then move on to the algorithm for general $A$, which is what we are using in the current implementation.

\paragraph*{Diagonal Case}
In the diagonal case, the integral separates into an x-, y- and z-dependent part. The total integral can thus easily be evaluated by using the explicit formula for the one-dimensional Gaussian integral, namely:
\begin{align}
    \langle x^{n_0}y^{n_1}z^{n_2}\rangle  & = \int_{\relc}\mathrm{d}^3r x^{n_1}y^{n_2}z^{n_3}e^{-\frac{1}{2}\left(\frac{x^2}{\sigma_1^2} + \frac{y^2}{\sigma_2^2} + \frac{z^2}{\sigma_3^2} \right)} \\
    & = 
    \left(2\sigma^2_1\right)^{n_1'}\Gamma(n_1')
    \left(2\sigma^2_2\right)^{n_2'}\Gamma(n_2')
    \left(2\sigma^2_3\right)^{n_3'}\Gamma(n_3').
\end{align}
An advantage of the diagonal formalism is that it can easily be adapted to non-Gaussian densities. We could, for instance, consider densities of the form
\begin{align}
    g_0(\br) = \exp\left[-\left(\frac{|x|}{\sigma_x}+\frac{|y|}{\sigma_y}+\frac{|z|}{\sigma_z}\right)\right]
\end{align}
This could be used to have densities that have the symmetry of a cube (if all $\sigma_j$ are equal) or other rectangular shapes. While integrating these in spherical coordinates would be a nightmare due to the absolute values, in Cartesian coordinates, the computation is relatively simple once the problem is brought into this diagonal form.
There could, therefore, be some value in having an implementation that works for the diagonalized case. It should be noted, however, that we are not directly working with the density. Instead, we are approximating the density with our basis functions. If the resolution of our basis functions is low, it might not be possible to properly distinguish Gaussians from rectangular densities. This should be kept in mind before spending too much time developing complicated schemes with minimal additional improvements.

\paragraph*{General Case}
Here, we use an iterative updating scheme. For later convenience, we define the covariance matrix $C = A^{-1}$ and work with it instead. To keep the notation closer to the implementation, we will index its components as $0,1,2$ rather than $1,2,3$, corresponding to the coordinate axes $x,y,z$. \\
We can initialize the first few moments, up to a global factor, by
\begin{align}
    \langle 1 \rangle & = \langle x^0y^0z^0\rangle = 1 \\
    \langle x \rangle & = a_x \\
    \langle y \rangle & = a_y \\
    \langle z \rangle & = a_z
\end{align}
We can compute all higher-order moments by using three recurrence relations\cite{nigam_recursive_2020}, one in the x-, y-, and z-directions, respectively. These are given by:\\
x-iteration:
\begin{align}
\langle x^{n_0+1}y^{n_1}z^{n_2}\rangle & = a_0 \langle x^{n_0}y^{n_1}z^{n_2}\rangle + C_{00}n_0 \langle x^{n_0-1}y^{n_1}z^{n_2}\rangle \\
 & + C_{01}n_1 \langle x^{n_0}y^{n_1-1}z^{n_2}\rangle 
   + C_{02}n_2 \langle x^{n_0}y^{n_1}z^{n_2-1}\rangle
\end{align}
y-iteration:
\begin{align}
\langle x^{n_0}y^{n_1+1}z^{n_2}\rangle & = a_1 \langle x^{n_0}y^{n_1}z^{n_2}\rangle + C_{10}n_0 \langle x^{n_0-1}y^{n_1}z^{n_2}\rangle \\
 & + C_{11}n_1 \langle x^{n_0}y^{n_1-1}z^{n_2}\rangle 
   + C_{12}n_2 \langle x^{n_0}y^{n_1}z^{n_2-1}\rangle
\end{align}
z-iteration:
\begin{align}
\langle x^{n_0}y^{n_1}z^{n_2+1}\rangle & = a_2 \langle x^{n_0}y^{n_1}z^{n_2}\rangle + C_{20}n_0 \langle x^{n_0-1}y^{n_1}z^{n_2}\rangle \\
 & + C_{21}n_1 \langle x^{n_0}y^{n_1-1}z^{n_2}\rangle 
   + C_{22}n_2 \langle x^{n_0}y^{n_1}z^{n_2-1}\rangle.
\end{align}
Warning: Please note that for any of the above iterations, it is possible that some exponents will become negative. A concrete example, in which we naively use the above formulae, would be the evaluation of $\langle x^2 \rangle$ from the $x$-iteration, which would give
\begin{align}
\langle x^2\rangle & = a_0 \langle x^1\rangle + C_{00}\cdot 1 \cdot \langle 1\rangle + C_{01}\cdot 0 \cdot \langle xy^{-1}\rangle + C_{02}\cdot 0 \cdot \langle xz^{-1}\rangle
\end{align}
In such cases, the terms containing negative exponents, strictly speaking, do not exist. As the multiplication by zero suggests, however, these terms do not contribute to the final result. The correct iteration in this case therefore is
\begin{align}
\langle x^2\rangle & = a_0 \langle x^1\rangle + C_{00}n_0 \langle 1\rangle.
\end{align}
In practice, depending on the specific implementation, it might not even be necessary to make special cases for this if all array elements are initialized to zero.
Please also note that the initializations only apply to normalized Gaussians that have a proper interpretation as a probability density. For other normalizations, the coefficients will all be multiplied by the same global factor. Thus, in practice, it makes sense to include this global factor in the initialization, since the iterative scheme will then automatically ensure that all higher order moments have the correct prefactor as well.

\subsection{Orthonormalization}
\label{si:normalization}
The coefficients $\langle nlm | \rho_i \rangle$ from the above procedures result from using non-orthonormal bases. While not strictly necessary, working in orthonormal bases ensures minimal overlap in mutual information between features\cite{goscinski_optimal_2021}. This discussion only applies to basis functions that are square integrable (e.g. GTO); non-square-integrable basis functions (e.g. monomial basis) cannot be orthonormalized. Below, we discuss how to scale our coefficients, taking the GTO basis as a representative example for any square-integrable basis.

We first choose the coefficients of our density expansion $\langle nlm | \rho_i \rangle$ to use the unnormalized GTO basis $R_{nl}(r) = r^{l+2n}e^{\frac{-r^2}{2\sigma^2}}$ for our radial expansion. The general unnormalized GTO $\phi_d = r^{d}e^{\frac{-r^2}{2\sigma_d^2}}$ has a finite square-integral over $\mathbb{R}^3$: $$I_d = \int_0^\infty {|\phi_d(r)|^2*r^2 dr} = 2^{-1}\sigma_d^{2d+3}*\Gamma(\frac{2d+3}{2})$$ and can hence be normalized with the constant $N_d = 1/\sqrt{I_d}$. In other words, our normalized GTO is $\Phi_d = N_d * \phi_d$, and $\int_0^\infty {|\Phi_d(r)|^2*r^2 dr} = 1$. After normalizing all the bases, we can take orthogonalize them using Löwdin Symmetric Orthonormalization\cite{lowdin_nonorthogonality_1970}. We first find the overlap matrix between two normalized GTOs: $$G_{ij} = \int_0^\infty \Phi_i \Phi_j r^2 dr.$$ The orthonormalization matrix is the inverse square root of the overlap matrix $G^{-1/2}$, which is guaranteed to exist because the overlap matrix is a gram matrix and is hence symmetric positive definite. Specifically, $G^{-1/2}$ is calculated by diagonalizing $G$, then taking the recipricol of the square root of the diagonal matrix: $G = MDM^T$, $G^{-1/2} = MD^{-1/2}M^T$. Then, we can apply this matrix to obtain a set of orthonormal GTO basis vectors $\hat{\Phi}_i$: $\hat{\Phi}_i = G^{-1/2}_{ij}\Phi_j$. We note that in our previous procedure, we could in theory construct an overlap matrix from unnormalized GTOs, but practically, calculating the overlap between unnormalized GTOs of high order, then inverting them to find the orthonormalization matrix, is not numerically stable.

\subsection{Tricks for Efficient Implementation}

\begin{figure}
\centering
\includegraphics[width=0.5\linewidth]{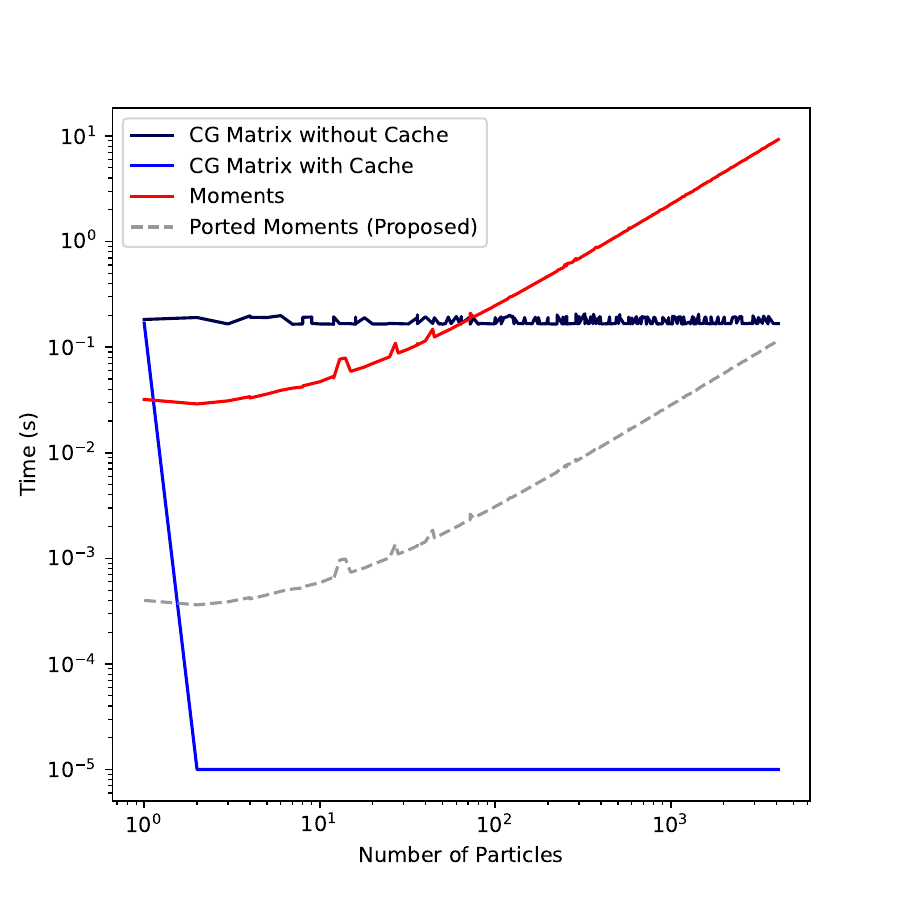}
\caption{\textbf{Time measurement of major bottlenecks} with and without caching. The same computation was performed for all N by N by M crystal structures with N and M both ranging from 1 to 16, and with fixed $l_{max} = 6, \sigma = 5.0, r_{cut} = 1.0$. With caching, after the first computation, all subsequent computations of Clebsch-Gordan matrices are near instantaneous ($\sim 10 \mu s$, the time required to access a list). The gray dashed line show the expected $\sim$80x speedup when we port the moments calculation to a lower-level language like Rust. The red and gray lines approach a slope of 1 on the log-log plot, indicating linear scaling with system size.}
\label{fig:benchmarks}
\end{figure}

The current implementation of \anisoap is performance limited, both by inefficient recalculations of the Clebsch-Gordan Matrices, and by the large number of computations and nested iterations performed in Python when calculating high-order moments. Below we outline two strategies under active development to address these inefficiencies. The computational costs of the current implementation and the proposed future implementations are shown in Fig.~\ref{fig:benchmarks}.

% We initially implemented \anisoap using a rudimentary evaluation, which was performance-limited by the computation of Clebsch-Gordan matrices to commute higher-order correlations and moments $\langle x^{n_0} y^{n_1} z^{n_2} \rangle$ inside the ellipsoidal transform. The timings for this rudimentary implementation are shown in Fig.~\ref{fig:benchmarks}.

\subsubsection{Caching the Clebsch-Gordan (CG) Matrix}
The construction of Clebsch-Gordan matrices depends on only the hyperparameter $l_{max}$. Originally, the matrices were stored within a Python dictionary indexed by $(l_1, l_2, L)$. Since $0 \leq l_1, l_2 \leq l_{max}$ and $|l_1 - l_2| \leq L \leq \min(l_{max}, (l_1 + l_2))$, the number of matrices required to compute grows as $\mathcal{O}(l_{max}^3)$, which is expensive even for modest values of $l_{max}$.

However, since the construction of CG matrices only depends on $l_{max}$, we can cache the matrices and re-use them internally. For caching, we use a simple cyclic list that stores \texttt{(key, value)} pair with \texttt{key} storing $l_{max}$ and \texttt{value} storing corresponding matrices, as shown below, where $M$ corresponds to the appropriate CG matrices.

\begin{tikzpicture}[every text node part/.style={align=center}]
\draw[-Stealth] (0.5, 1.5) -- (0.5, 1);

\draw (0,0) rectangle (1,1);
\node at (0.5,0.5) {$key_{1}$ \\ $M_{1}$};
\draw (1,0) rectangle (2,1);
\node at (1.5,0.5) {$key_{2}$ \\ $M_{2}$};
\node at (2.5,0.5) {$\cdots$};
\draw (3,0) rectangle (4.5,1);
\node at (3.75,0.5) {$key_{n - 1}$ \\ $M_{n - 1}$};
\draw (4.5,0) rectangle (6,1);
\node at (5.25,0.5) {$key_{n}$ \\ $M_{n}$};
\end{tikzpicture}

The cache has a finite and fixed number of \texttt{(key, value)} pairs it can store to prevent memory overflow, and whenever the cache is full, it decides the entry for replacement using an algorithm that mimics the CLOCK algorithm for page replacement\cite{corbato_paging_1969}. In accordance with the algorithm, the \texttt{key} for the CG list also contains a replacement bit to store the time-indexed usage of the entry. The format of the CLOCK algorithm does place implicit limits on $l_{max}$, such that $l_{max} \leq 2^{31} - 1$, although this value is beyond practical usage. The utility of caching these CG matrices is apparent when performing repeated higher-body order expansions with the same $l_{max}$, such as in a molecular dynamics simulation.

\subsubsection{Moments Array}
Given the simplicity of the moments calculations and the frequency of invoking this section of code, we are currently porting this functionality to Rust\cite{matsakis_rust_2014}, which gives the benefits of a compiled language while retaining easy interfacing with Python. 

While the Rust code is almost a one-to-one translation, there were two minor changes. Firstly, the computation of the inverse of dilation matrix $\bD$ \eqref{eq:dilmat} is changed. By definition, $\bD$ is a 3 by 3 symmetric matrix. Therefore, in Rust we compute the analytical formula for inverting this matrix to avoid unnecessary operations while minimizing package dependency. Secondly, the code was re-organized in Rust to maintain clarity in the new syntax.

\subsubsection{Results of Optimization}
With the CG-matrix caching and proposed porting of the moments above, we have performed benchmark timings (Fig.~\ref{fig:benchmarks}) to compare the speed of execution. These benchmarks were computed on the High-Performance Computers (HPC) within the Wisconsin Center for High-Throughput Computation. Each HPC had a 3GHz AMD EPYC 7763 64-Core Processor and 514 GB Total RAM. Note that none of this code is currently parallelized and therefore only a single core was used, yielding timings that are similar to what is obtained on a local machine. 

\section{Gay-Berne Case-Study}
\subsection{Gay-Berne Interactions}
\label{si:gb}
Continuing our discussion of Eq.~\eqref{eq:gb}, the second term, $\eta_{12}$, is a function of each ellipsoids' geometry and orientation but not a function of distance, and is given as follows:

\begin{equation}
    \eta_{12}(\rot_1, \bS_1, \rot_2, \bS_2) = \left[\frac{2s_1s_2}{\det\boldsymbol{\Xi}_{12}(\rot_1, \rot_2)]}\right]^{\nu/2}
\end{equation}
\begin{equation}
    s_i = [a_i b_i + c_i c_i][a_i b_i]^{1/2}
\end{equation}

This correction term describes the interaction strength between two ellipsoids at 0 separation, whose influence is tuned by parameter $\nu$. %\al{A Gay–Berne potential for dissimilar biaxial particles, by BFZ}.

The last term, $\chi_{12}$, is only a function of each ellipsoid's individual orientations ($\rot_1, \rot_2$) and relative orientation ($\hat{r_{12}}$), but not of their geometries ($\bS_1, \bS_2$). It is given as follows:
\begin{equation}
    \chi_{12}(\rot_1, \rot_2, \hat{r}_{12}) = \left[\hat{r}_{12}^T \boldsymbol{B}_{12}^{-1}(\rot_1,\rot_2) \hat{r}_{12} \right]^\mu
\end{equation}
\begin{equation}
    \boldsymbol{B}_{12}(\rot_1,\rot_2) = \rot_1^T\boldsymbol{E}_1\rot_1 + \rot_2^T\boldsymbol{E}_2\rot_2
\end{equation}
\begin{equation}
    \boldsymbol{E}_i = \begin{pmatrix}
        &e_{ai}^{-1/\mu} &0 &0 \\ 
        &0 &e_{bi}^{-1/\mu} &0 \\
        &0 &0 &e_{ci}^{-1/\mu} \\ 
    \end{pmatrix}
\end{equation}

Note that for Gay-Berne, the rotations $\rot_i$ transform from lab frame to the body frame, while \anisoap defines rotations to transform from the body frame to the lab frame. Hence, $\rot_{i,\AniSOAP} \equiv \rot_{i,GB}^{-1}$. The above definitions use $\rot_i \equiv \rot_{i, GB}$.

$\chi_{12}$ corrects for the well-depth by interpolating the relative well depth between the side-to-side, face-to-face, and end-to-end interactions, given by $e_{ai}$, $e_{bi}$, $e_{ci}$. Generally, these pole-pole relative well depths are arbitrarily fitted, but can be assumed to be equal to the gaussian curvature of the ellipsoids at each pole, provided that $\mu=1$:

\begin{equation}
    \boldsymbol{E}_i = \sigma \begin{pmatrix}
        &\frac{a_i}{b_ic_i} &0 &0 \\ 
        &0 &\frac{b_i}{a_ic_i} &0 \\
        &0 &0 &\frac{c_i}{a_ib_i} \\ 
    \end{pmatrix}
    \label{eq:gb-well-depths2}
\end{equation}

In our case-study with $a_i=1, b_i=1.5, c_i=2$ ellipsoids, we set $\mu=1$, enabling the use of \ref{eq:gb-well-depths2} to calculate $\boldsymbol{E}_i$. We furthermore set $\nu=1$, completely specifying the Gay-Berne hyperparameters.
\clearpage

\section{Supplementary Information}
\subsection{Liquid Crystals}
\anisoap is well-suited for datasets containing multiple ellipsoid types; however, it is not necessarily a fair comparison with the established order parameters to demonstrate their inability to reconstruct such complex datasets. Here we include additional analyses for identical frames as those used in Sec.~\ref{sec:lc}, augmenting by another 1000 frames that represent $L/D=2$ spheroids.

\twocolumngrid

\begin{figure}[h!]
    \centering
    \includegraphics[width=0.9\linewidth]{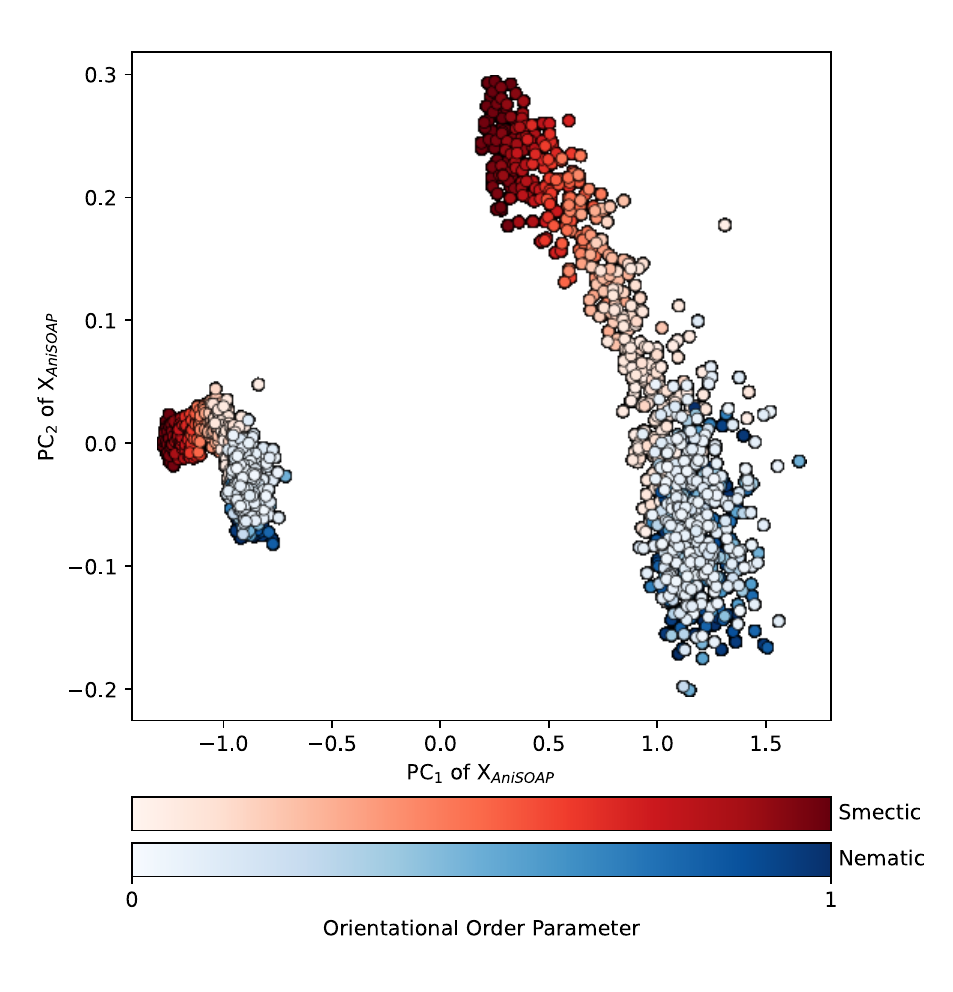}
    \caption{Analogous PCA to Fig.~\ref{fig:lc_pca}, with both L/D=3 (left) and L/D=2 (right) spheroids.}
    \label{si:lc_pca_hetero}
\end{figure}

\begin{figure}[h!]
    \centering
    \includegraphics[width=0.9\linewidth]{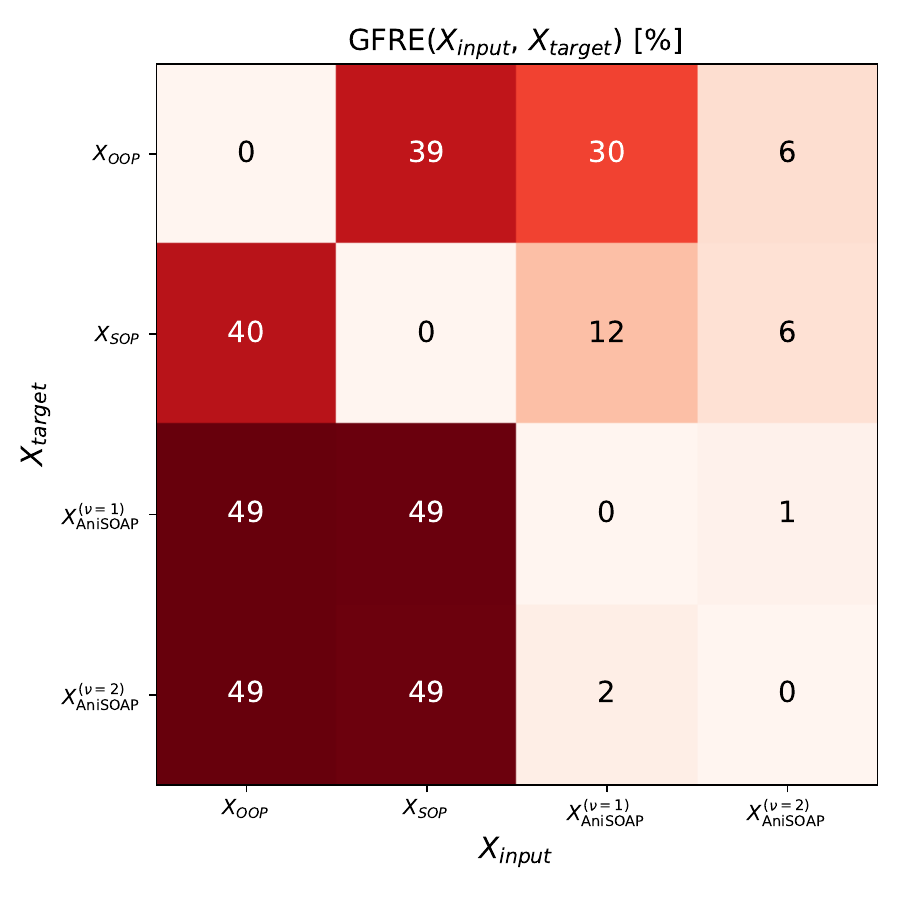}
    \caption{Analogous to Fig.~\ref{fig:lc_gfre}, the reconstruction of different representations of our augmented dataset.}
    \label{si:lc_gfre}
\end{figure}

\onecolumngrid

Our GFRE results emphasize that \anisoap is able to contain analogous information to both the Steinhardt and Orientational Order Parameters; however, the reconstruction of \anisoap decreases by either representation when our dataset contains multiple particle types.

\begin{figure}[!h]
    \centering
    \includegraphics[width=0.8\linewidth]{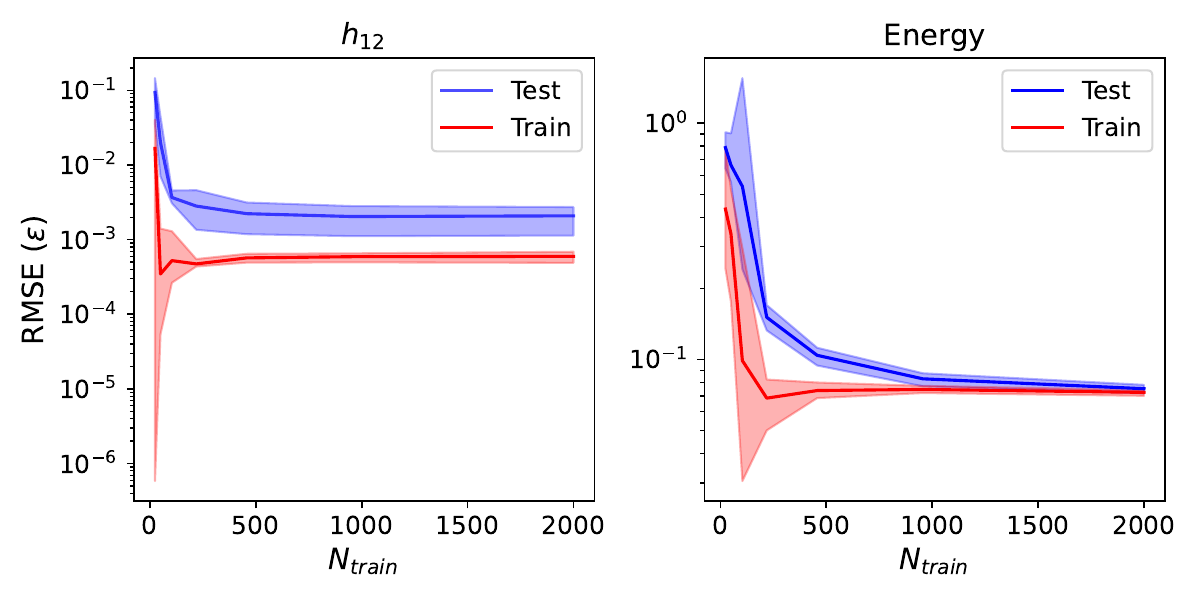}
    \caption{Learning curve from Gay-Berne dataset, showing RMSE values for different training set sizes. The solid line is the average RMSE over 5 folds, while the minimum and maximum RMSE denoted by the boundaries of the shaded region. Note that $N_{train}=2000$ corresponds with a training size consistent with 10\% of our full dataset.}
    \label{si:gb_learning_curve}
\end{figure}

\subsection{Gay-Berne Learning Curve}
We computed the learning curves for the Gay-Berne dataset contained in the main text using the \texttt{learning\_curve} functionality in \texttt{scikit-learn}, assuming cross-validated ridge regression as our underlying estimator, five-fold cross-validation, and RMSE as the scoring mechanism. The learning curves for the Gay-Berne dataset demonstrate that even small train-test splits can be used to train a highly-accurate ridge-regression predictor on the \anisoap representation, showing its powerful interpolation abilities on both $h_{12}$ and energies.

\subsection{Benzene Crystals Case-Study}
\subsubsection{Learning Curve}

We computed learning curves for the dataset contained in the main text using the \texttt{scikit-learn} learning curve functionality, assuming cross-validated ridge regression as our underlying estimator, five-fold cross-validation, and RMSE as the scoring mechanism. Our plot shows that we are within the saturation regime with the learning of this dataset using the \anisoap representation. This learning curve also demonstrates that \anisoap can effectively learn benzene energetics within minimal training data.

\begin{figure}[!h]
    \centering
    \includegraphics[width=0.8\linewidth]{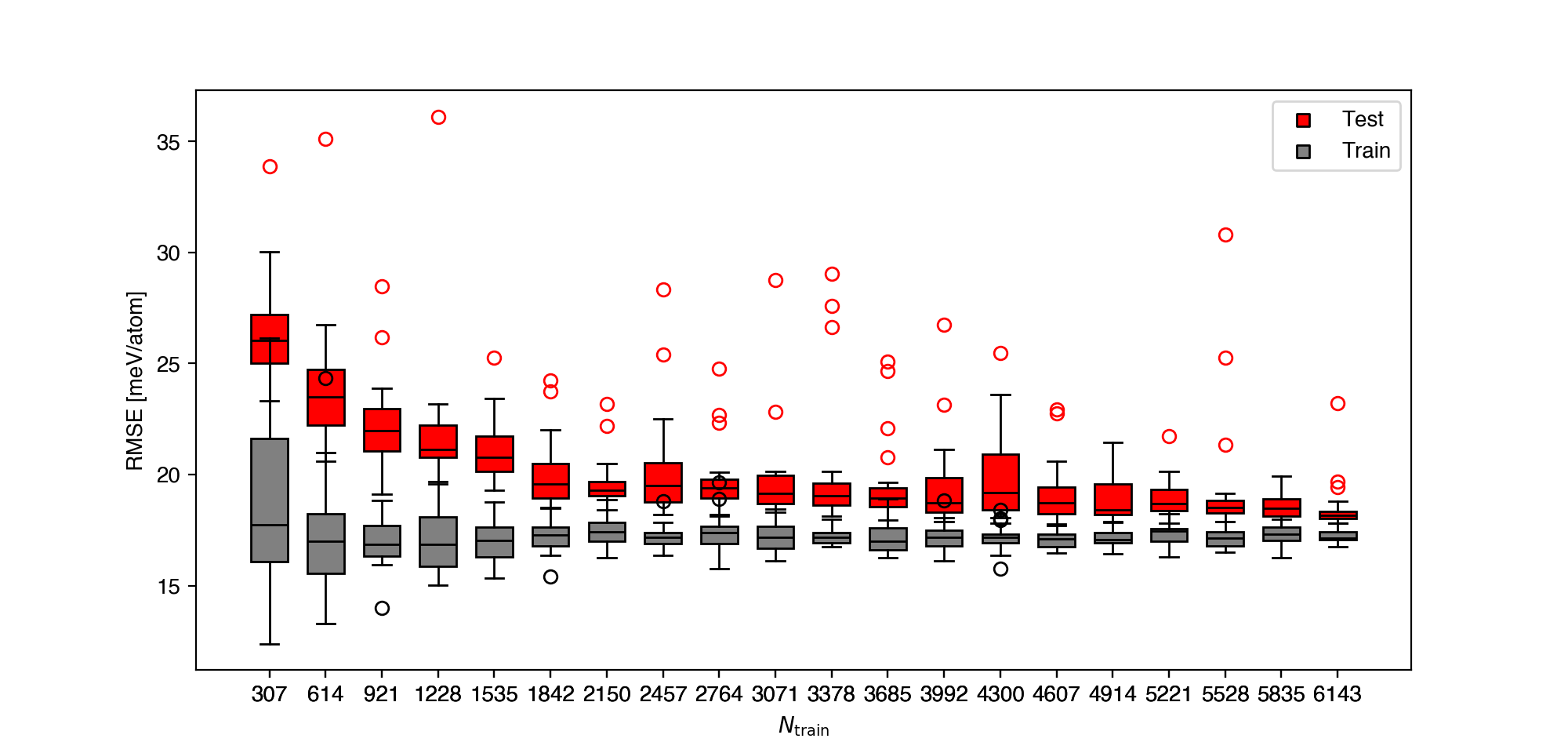}
    \caption{Learning curve for benzene dataset, showing root-mean-squared-error values for different training set sizes. Circles denote outliers in the learning curve, as determined by those falling outside the second quartile of the distribution by \texttt{matplotlib.pyplot.boxplot}.}
    \label{si:benzene_learning}
\end{figure}

\subsubsection{Hyperparameter Tuning: Additional Results}

\begin{figure}[!h]
    \centering
    \includegraphics[width=0.8\linewidth]{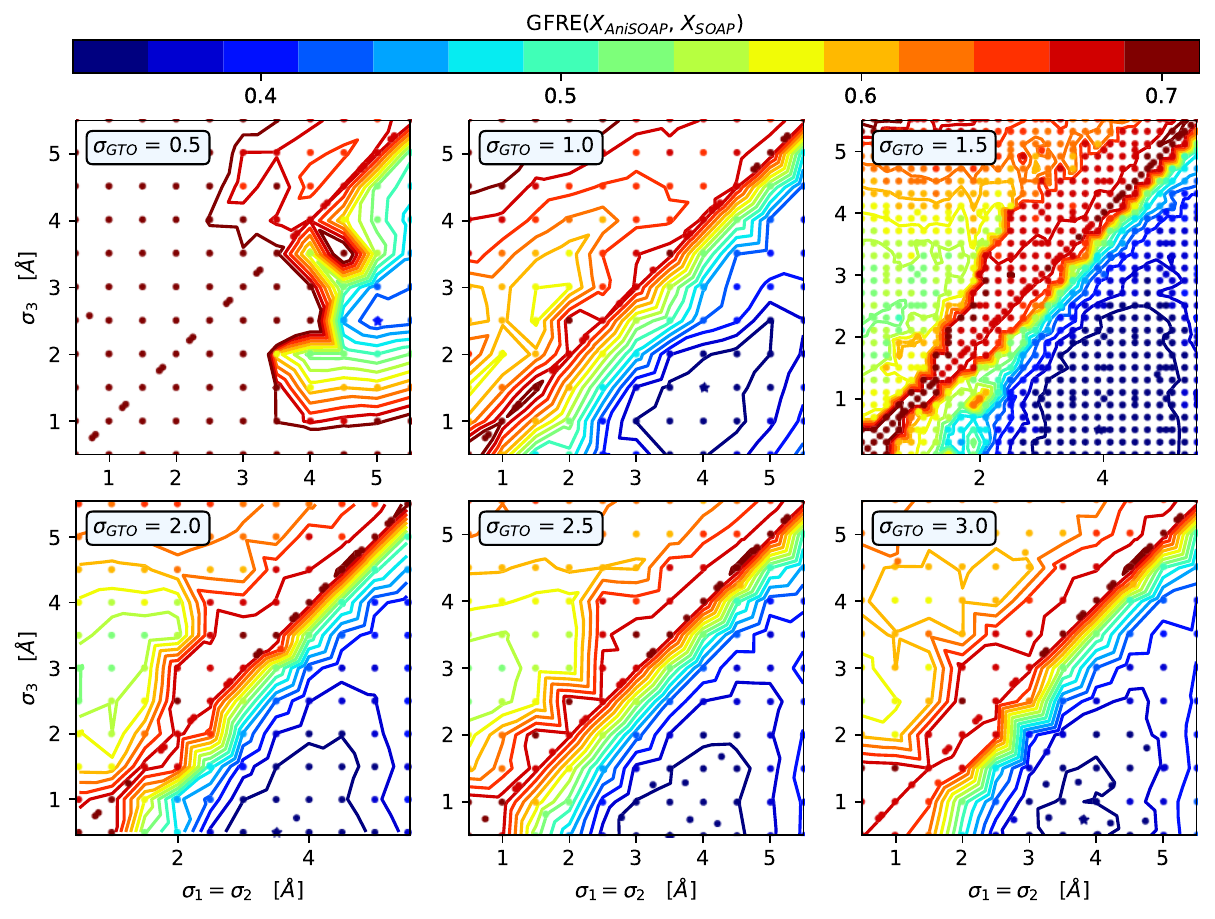}
    \caption{Additional results from hyperparameter running for $\sigma_\text{GTO}=0.5, 1.0, 1.5, 2.0, 2.5, 3.0$\r{A}. Color of the scatter and contour lines denote $GFRE(X_\text{AniSOAP}, X_\text{SOAP})$, where contour levels are consistent across all plots and in the main text. For $\sigma_{GTO}=0.5$, many points at $\sigma_{1,2}<3.0$ had results greater than 0.7; however, we truncated the color map to be consistent with the main text and show the gradation of the oblate region more clearly.}
    \label{fig:si_gfre}
\end{figure}

\end{document}

%% file: defs.tex
\newcommand{\rot}{\mathcal{R}}
\newcommand{\wigD}{\mathcal{D}}

\newcommand{\bx}{\boldsymbol{x}}
\newcommand{\bX}{\boldsymbol{X}}

\newcommand{\br}{\boldsymbol{r}}
\newcommand{\brij}{\br_{ij}}

\newcommand{\ba}{\boldsymbol{a}}

\newcommand{\bA}{\mathbf{A}}
\newcommand{\bD}{\boldsymbol{D}}
\newcommand{\bS}{\boldsymbol{S}}

\newcommand{\real}{\mathbb{R}}
\newcommand{\relc}{\mathbb{R}^3}

\newcommand{\gij}[1]{
\ifthenelse { \equal {#1} {} }  
    {g(\br - \brij)}
    {g(#1)}
}
\newcommand{\ylm}[1]{
\ifthenelse { \equal {#1} {} }  
    {Y_l^m(\hat{r})}
    {Y_l^m(#1)}
}
\newcommand{\ylmp}[1]{
\ifthenelse { \equal {#1} {} }  
    {Y_l^{m\prime}(\hat{r})}
    {Y_l^{m\prime}(#1)}
}
\newcommand{\rnl}[1]{
\ifthenelse { \equal {#1} {} }  
    {R_{nl}(r)}
    {R_{nl}(#1)}
}
\newcommand{\rlm}[1]{
\ifthenelse { \equal {#1} {} }  
    {R_{l}^{m}(\hat{r})}
    {R_{l}^{m}(#1)}
}
\newcommand{\pnlm}{\braket{nlm | \rho_{ij}}}

\newcommand{\eg}{e.g.,~}

\usepackage{xcolor}

\newcommand{\AniSOAP}{{AniSOAP}~}
\newcommand{\anisoap}{\AniSOAP}